\documentclass[aps,prd,showpacs,twocolumn,
superscriptaddress]{revtex4-1}
\usepackage{tabularx}
\usepackage{multirow}
\usepackage{graphicx}
\usepackage{dcolumn}
\usepackage{float}
\usepackage{morefloats}
\usepackage{amsmath}
\usepackage{amssymb}
\usepackage{bm}
\usepackage{color}
\usepackage[normalem]{ulem}
\usepackage[dvipsnames]{xcolor}
\usepackage{hyperref}
\usepackage{placeins}

\hypersetup{
colorlinks=true, 
linkcolor=blue,  
citecolor=cyan,  
}

\begin{document}
\title{Particle motion around generic black holes coupled to non-linear electrodynamics}

\author{Jaroslav Vrba}
\email{jaroslav.vrba@fpf.slu.cz}
\affiliation{Institute of Physics and Research Centre of Theoretical Physics and Astrophysics, Faculty of Philosophy \& Science, Silesian University in Opava,
Bezru\v{c}ovo n\'{a}m\v{e}st\'{i} 13, CZ-74601 Opava, Czech Republic}

\author{Ahmadjon Abdujabbarov}
\email{ahmadjon@astrin.uz}
\affiliation{Amity University in Tashkent, 70 Labzak street, 100028, Tashkent, Uzbekistan}
\affiliation{Ulugh Beg Astronomical Institute, Astronomicheskaya 33,
	Tashkent 100052, Uzbekistan }
\affiliation{National University of Uzbekistan, Tashkent 100174, Uzbekistan}

\author{Arman~Tursunov}
\email{arman.tursunov@fpf.slu.cz}
\affiliation{Institute of Physics and Research Centre of Theoretical Physics and Astrophysics, Faculty of Philosophy \& Science, Silesian University in Opava,
Bezru\v{c}ovo n\'{a}m\v{e}st\'{i} 13, CZ-74601 Opava, Czech Republic}
\affiliation{Ulugh Beg Astronomical Institute, Astronomicheskaya 33,
	Tashkent 100052, Uzbekistan }
\affiliation{I. Physikalisches Institut der Universität zu Köln, Zülpicher Strasse 77, 50937 Cologne, Germany}

\author{Bobomurat Ahmedov}
\email{ahmedov@astrin.uz}
\affiliation{Ulugh Beg Astronomical Institute, Astronomicheskaya 33,
	Tashkent 100052, Uzbekistan }
\affiliation{National University of Uzbekistan, Tashkent 100174, Uzbekistan}

\author{Zden\v{e}k Stuchl\'{i}k}
\email{zdenek.stuchlik@fpf.slu.cz}
\affiliation{Institute of Physics and Research Centre of Theoretical Physics and Astrophysics, Faculty of Philosophy \& Science, Silesian University in Opava,
Bezru\v{c}ovo n\'{a}m\v{e}st\'{i} 13, CZ-74601 Opava, Czech Republic}

\date{\today}
\begin{abstract}
We study spherically symmetric magnetically charged generic black hole solutions of general relativity coupled to non-linear electrodynamics (NED). For characteristic values of the generic spacetime parameters we give the position of horizons in dependence on the charge parameter, demonstrating separation of the black hole and no-horizon solutions, and possibility of existence of solutions containing three horizons. We show that null, weak and strong energy conditions are violated when the outer horizon is approaching the center. We study effective potentials for photons and massive test particles and location of circular photon orbits (CPO) and innermost stable circular orbit (ISCO). We show that the unstable photon orbit can become stable, leading to the possibility of photon capture which affects on silhouette of the central object. The position of ISCO approaches the horizon with increasing charge parameter $q$ and the energy at ISCO decreases with increasing charge parameter. We investigate this phenomenon and summarize for a variety of the generic spacetime parameters the upper estimate on the spin parameter of the Kerr black which can be mimicked by the generic charged black hole solutions.
\end{abstract}

\pacs{04.50.-h, 04.40.Dg, 97.60.Gb}

\maketitle

\section{Introduction}
Physical singularity at the center of the standard vacuum solutions within the Einstein general relativity exists due to the Penrose and Hawking theorems~\cite{Hawking75}. However, there are several effective methods to avoid the singularity, and get a singularity-free non-vacuum or modified gravity solution describing compact objects in astrophysics. The conformal gravity models~\cite{Bambi17, Bambi17d, Toshmatov17b, Chakrabarty18a,Bambi18}, quantum corrections~\cite{Horava09b,Horava10}, dark energy star - gravastar models~\cite{Mazur2004, Chirenti07, Turimov09a, Chirenti16} can be considered as examples of models providing the singularity-free solutions.  

Other interesting way of avoiding the singularity is  solution of general relativity coupled to non-linear electrodynamics. Recently the general procedure for obtaining the solutions describing the compact objects with the electric and/or magnetic charges in general relativity coupled to non-linear electrodynamics has been described~\cite{Fan16,Toshmatov18b}. Such generic solutions could be quite regular black hole (or no-horizon) solutions, if their mass parameter is related exclusively to the NED part~\cite{Fan16,Toshmatov18b}. In Ref.~\cite{Toshmatov18b} authors suggested the regular black hole models as the gravitational field of a non-linear electric or magnetic monopole. The properties of the rotating analogue of this type black hole have been considered in~\cite{Toshmatov17d}. In the special cases of generic regular NED black holes and no horizon spacetimes (Bardeen, Hayward) the motion of particles and photons was investigated in~\cite{StuSche15,ScheStu15,AS14,HZSD18} 

Here we study the properties of the singular generic black hole solutions containing also the standard self-interacting mass parameter, analyze the energy conditions and test particle motion in their environment.  

The main purpose of the studying the particle motion around compact objects is to test the model or solution of gravity theories. Particularly, the accretion disc around the astrophysical black holes is considered as a relevant model of some X-ray sources and the observation of inner edge radii of accretion disc can give constraints on parameters of alternative and modified theories of gravity~\cite{Tursunov16,Abdujabbarov08, Abdujabbarov10,Abdujabbarov11a, Abdujabbarov13a,Abdujabbarov13b,Abdujabbarov14,Stuchlik14a, Frolov10,Frolov12,Karas12a,Stuchlik16, Kovar10,Kovar14,Kolos17, Tripathi19,Bambi16b, Cao18,Kong14, Bambi17e}. 
Above presented facts are good motivation for study test particle's motion around black hole described by the general relativistic solution coupled with non-linear electrodynamics.  

The paper is organized as follows. In Sect.~\ref{2gbhcondi} we describe generic black hole and consider existence of horizons and at energy conditions.  Sect.~\ref{3circ} is devoted to studying the motion and circular orbits of photons and neutral massive test particles around generic black hole. In Sect.~\ref{4Mimic} the possibility to distinguish the spin  of the Kerr black hole from the charge parameter for the generic black hole solution coupled to NED is discussed. Finally, in Sect.~\ref{5Summary} we summarize our obtained results. 

Throughout the paper we use a space-like signature $(-,+,+,+)$, a system of units in which $G = c = 1$ and we restore them when we need to compare our results with observational data. Greek indices run from $0$ to $3$, Latin indices from $1$ to $3$.

\section{Generic black hole horizons and energy conditions \label{2gbhcondi}}
\subsection{Generic black hole spacetime}
The generic black hole spacetime metric has the following Schwarzschild-like form in the standard Schwarzschild coordinates 
\begin{eqnarray}\label{bhmetric}
ds^2&=&- \left(1-\frac{2 m(r)}{r}\right)dt^2+ \left(1-\frac{2 m(r)}{r}\right)^{-1} dr^2\nonumber\\
&& +r^2 d\theta^2 + r^2\sin^2\theta d\phi^2.
\end{eqnarray}
We consider the family of solutions where the mass function $m(r)$ is defined as~\cite{Fan16,Toshmatov18b}
\begin{eqnarray}\label{massf1}
m(r)&=& M-\frac{q^3}{\alpha}\Big[1-\frac{r^\mu}{(r^\nu+q^\nu)^{\mu/\nu}}\Big],
\end{eqnarray}
where $M$ is the pure gravitational self-interaction mass and the Arnowitt-Deser-Misner (ADM) mass is defined as 
$$
M_{\rm ADM}= M+M_{\rm em},
$$ 
with 
$$
M_{\rm em} = \frac{q^3}{\alpha}
$$
being the electromagnetically induced gravitational mass, where $\alpha>0$ describes the strength of the non-linear effect, $\mu>0$ characterizes the degree of non-linearity, $q$ is the integration constant related to the magnetic charge by relation \ref{Qqrelation} and $\nu>0$ is free parameter governing character of its field in concrete solution and we suppose $q>0$ as in~\cite{FW16}. 
The Lagrangian density governing the electromagnetic non-linear self-interaction takes the form~\cite{FW16}
\begin{eqnarray}
\mathcal{L}=\frac{4\mu}{\alpha}\frac{(\alpha F)^{\frac{\nu+3}{4}}}{[1+(\alpha F)^{\frac{\nu}{4}}]^{1+\frac{\mu}{\nu}}}, \nonumber
\end{eqnarray}
where $F=F^{\mu\nu}F_{\mu\nu}$.
The metric (\ref{bhmetric}) with mass function (\ref{massf1}) is used throughout this paper.

The magnetic charge can be expressed as~\cite{Fan16,Toshmatov18b,Toshmatov17d}
\begin{eqnarray}\label{Qqrelation}
Q_m=\frac{q^2}{\sqrt{2 \alpha}}.
\end{eqnarray}

The regular black hole (no physical singularities even at the center $r=0$) appears when we assume
\begin{eqnarray}\label{meqmem}
M = M_{\rm em} ,
\end{eqnarray} 
with condition $\mu\geq 3$~\cite{FW16}. Requirement (\ref{meqmem}) with (\ref{massf1}) leads to mass function 
\begin{eqnarray}\
m(r)=\frac{Mr^\mu}{(r^\nu+q^\nu)^{\mu/\nu}}.
\end{eqnarray}
The electromagnetic field potential of the non-rotating magnetically charged black hole spacetime takes the form
\begin{eqnarray}
A_\alpha = \left(0,\ 0,\ 0,\ Q_m \cos\theta  \right).
\end{eqnarray}

The magnetic charge in a black hole could be a remnant of the phase transitions in the early Universe where in accord with Grand-Unification theory magnetic monopoles could be created as defects of the transition \cite{Kolb90}. Alternatively see~\cite{Kruglov19a,Kruglov18,Kruglov16b}.

In the following we consider the generic singular black hole solutions with the self-gravitating term $M\neq0$, i.e., we do not assume condition (4) for the non-singular black holes in the rest of the paper. We can see immediately that for $r\rightarrow \infty$ there is $m(r)\rightarrow M$, and the mass parameter $M$ governs the gravitational mass at large distances. We can thus define dimensionless radius $r/M\rightarrow r$ and dimensionless charge parameter $q/M\rightarrow q$. Then we can use dimensionless parameter $\alpha/M^2 \rightarrow \alpha$. For using of these dimensionless quantities for simplicity we can put $M=1$.

\subsection{Generic black hole horizons}

Now we explore existence and behavior of the horizon(s) of generic black holes. The lapse function can be written in the form
\begin{eqnarray}\label{lapsf}
f(r)&=&-g_{tt}(r)=1-\frac{2m(r)}{r}\\ \nonumber
&=&1-\frac{2}{r}\Bigg\{1-\frac{q^3}{\alpha}\Big[1-\frac{r^\mu}{(r^\nu+q^\nu)^{\mu/\nu}}\Big] \Bigg\}.
\end{eqnarray}

\begin{figure*}
\includegraphics[width=1\hsize]{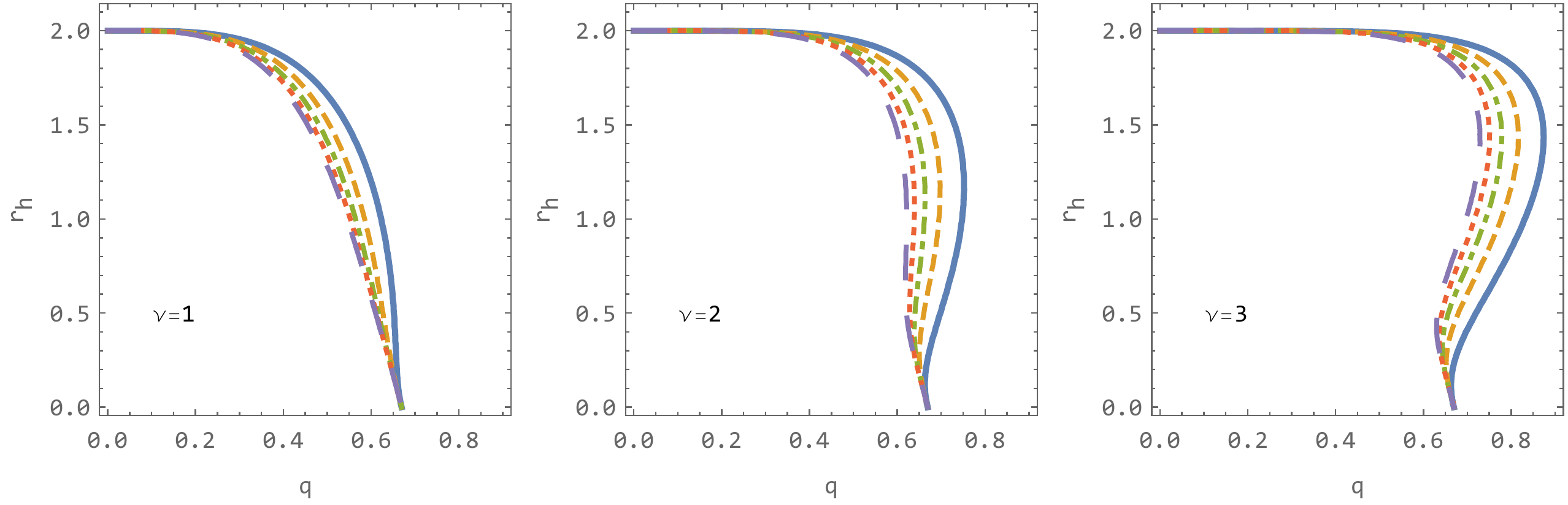}
\caption{\label{hornu}The location of horizon of generic black hole spacetime with parameters $\alpha=0.3$, $\nu=1,2,3$ and values $\mu$: $\mu=2$ (\textit{blue}), $\mu=3$ (\textit{orange}), $\mu=4$ (\textit{green}), $\mu=5$ (\textit{red}) and $\mu=6$ (\textit{purple}).}
\end{figure*}

\begin{figure*}
\includegraphics[width=1\hsize]{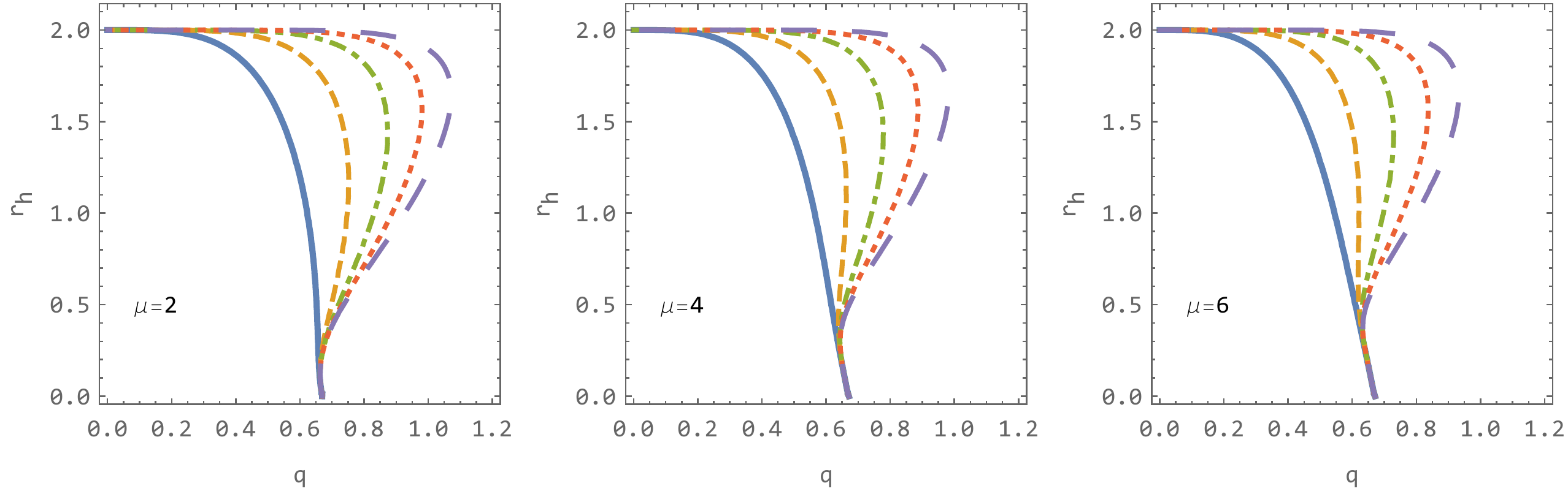}
\caption{\label{hormu}The location of horizon of generic black hole spacetime with parameters $\alpha=0.3$, $\mu=2,4,6$ and values $\nu$: $\nu=1$ (\textit{blue}), $\nu=2$ (\textit{orange}), $\nu=3$ (\textit{green}), $\nu=4$ (\textit{red}) and $\nu=5$ (\textit{purple}).}
\end{figure*}

\begin{figure}
\includegraphics[width=1\hsize]{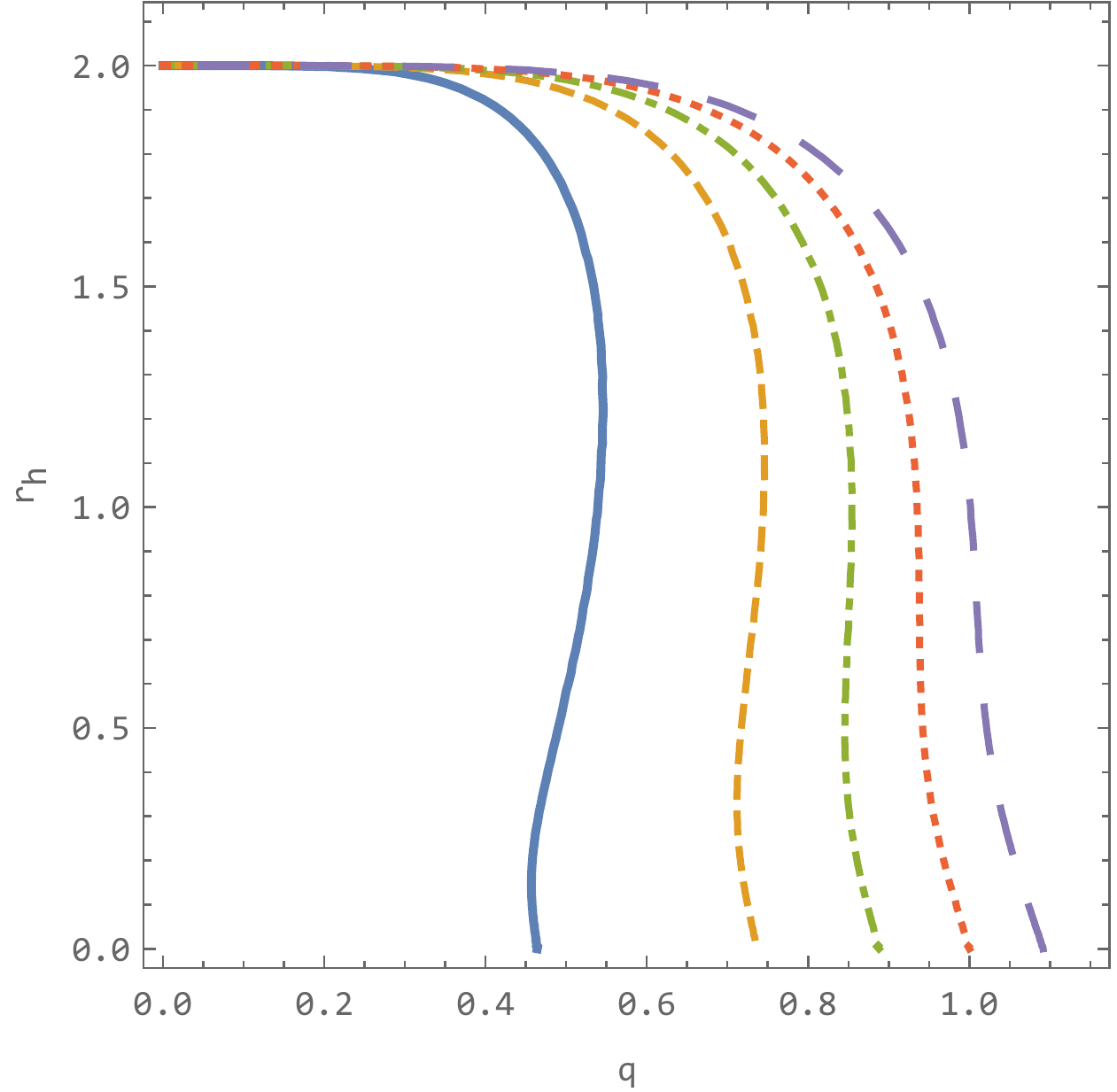}
\caption{\label{hora}The location of horizon of generic black hole spacetime with parameters $\mu=3$, $\nu=2$ and values $\alpha$: $\alpha=0.1$ (\textit{blue}), $\alpha=0.4$ (\textit{orange}), $\alpha=0.7$ (\textit{green}), $\alpha=1.$ (\textit{red}) and $\alpha=1.3$ (\textit{purple}).}
\end{figure}

 Four families can be found depending on the spacetime parameters. One can distinguish them according to the number of horizons from zero to three possible horizons which can appear in the considered generic black hole spacetime.
 The existence and location of horizons comes from a solution of equation $g_{tt}=0$, it follows from (\ref{lapsf}) that we have to solve equation
 \begin{eqnarray}\label{horeq}
 2q^3\Big( r^\mu(r^\nu+q^\nu)^{-\mu/\nu} -1 \Big)=\alpha\big( r-2 \big).
 \end{eqnarray}
However, Eq. (\ref{horeq}) cannot be solved analytically - the results of numerical calculations are illustrated in Figs. \ref{hornu} - \ref{hora}. 
Illustration of the behavior of the horizon for a variety of $\mu$ and $\nu$ parameters is shown in Figs. \ref{hornu} and \ref{hormu}.
Interesting characteristics is related to the existence of limiting value of charge parameter $q_L$ for which all horizons disappear and the object becomes a naked singularity for any bigger value $q>q_L$. We can see that $q_L$ depends on all the other black hole parameters, the greater is parameter $\nu$ the greater is $q_L$. The parameter $\mu$ has evidently weaker and opposite effect. Influence of the parameter $\alpha$ on horizon is plotted in Fig. $\ref{hora}$ and shows how $q_L$ increases with parameter $\alpha$. The outer horizon also increases with increasing of the parameter $\alpha$ (for fixed $q$), since the negative electromagnetic contribution of mass decreases.

Very interesting result appears when the "innermost" horizon shrinks to zero. At this point the object can lose physical singularity and become non-singular (if $\mu\geq 3$). Eq. (\ref{meqmem}) then implies $q^3=\alpha$. For that case there may exist two, one or zero horizon.
In case of non-singular compact object, for the chosen values of parameters $\mu$ and $\nu$, the number of horizons can range from one to three, while one of the horizons is always located at $r=0$, as can be seen e.g. in Fig.\ref{hornu}. Existence of the event horizon defines the existence of black hole. Therefore the object having zero-radius horizon and two more horizons is a non-singular black hole. The outermost horizon is the event horizon, the middle horizon is a form of the Cauchy horizon. When the two horizons coincide, the object is a non-singular extremal black hole with zero-radius horizon and event horizon.

\subsection{Energy conditions}\label{enconss}

Here we study energy conditions for spacetime described by (\ref{bhmetric}) and (\ref{massf1}), using the Einstein equations  $G^\mu_\nu=T^\mu_\nu$, where $T^\mu_\nu={\rm diag}(-\rho,P_1,P_2,P_3)$ is the stress-energy tensor releted  to electromagnetic field. It is also obvious (because of the symmetry of spacetime) that $P_2=P_3$. We assume the above assumptions on parameters and black hole solution for the rest of this paper.  The energy conditions were tested in \cite{Toshmatov17b} or \cite{Denisov17,Kruglov16a,Kruglov19b} and even more restrictive causality and unitarity conditions were studied in \cite{Shabad11}.
\subsubsection{Null energy condition}
The null energy condition (NEC) can be written as
\begin{eqnarray}
T_{\mu\nu} n^\mu n^\nu \geq 0,
\end{eqnarray}
where $n^\mu$ is a null-like vector. We can rewrite it and show the physical meaning in the form
\begin{eqnarray}
\rho + P_i \geq 0,\quad i = 1,2,3.
\end{eqnarray}

The symmetry of the spacetime ($G^t_t=G^r_r$) helps again because for $i=1$ it is identity. We study only $\rho + P_2 \geq 0$ for now and the NEC takes the form 
\begin{eqnarray}\label{enec}
(\nu +3) r^{\nu }-(\mu -3) q^{\nu }\geq 0.
\end{eqnarray}

\begin{figure*}
\includegraphics[width=1\hsize]{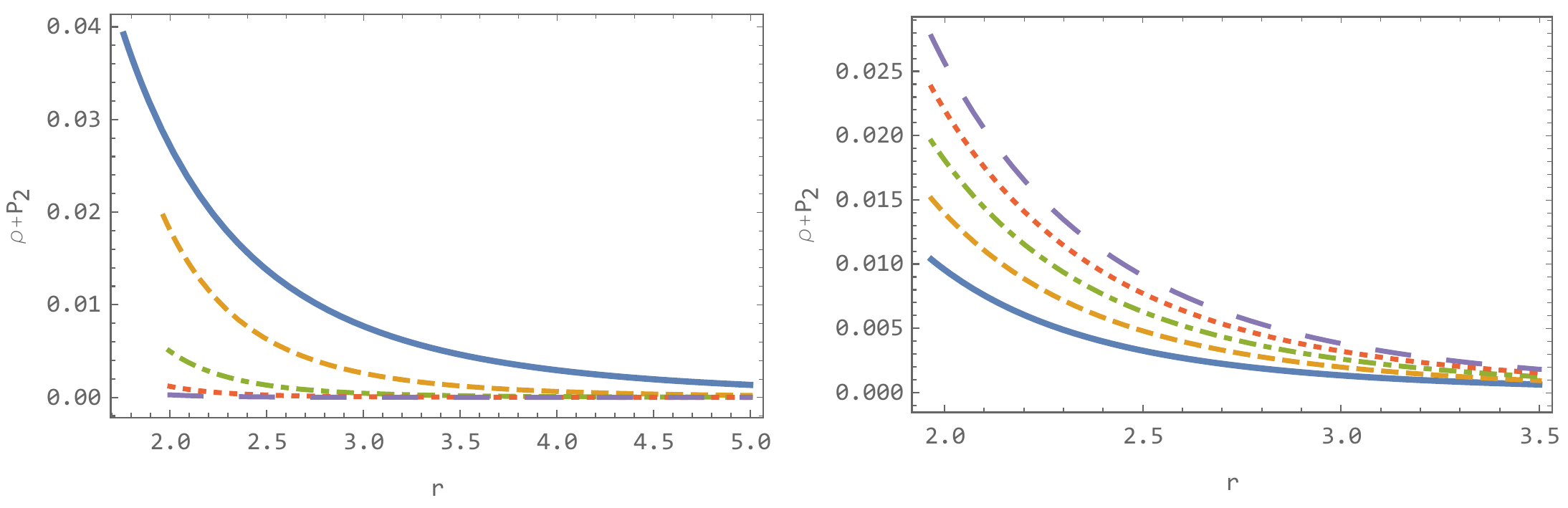}
\caption{\label{fNEC}NEC ($\rho + P_2$) of generic black hole spacetime and parameters $q=0.4$ and $\alpha=0.3$. \textit{Left panel} $\mu=4$ and values $\nu$: $\nu=1$ (\textit{blue}), $\nu=2$ (\textit{orange}), $\nu=3$ (\textit{green}), $\nu=4$ (\textit{red}) and $\nu=5$ (\textit{purple}). \textit{Right panel} $\nu=2$ and values $\mu$: $\mu=2$ (\textit{blue}), $\mu=3$ (\textit{orange}), $\mu=4$ (\textit{green}), $\mu=5$ (\textit{red}) and $\mu=6$ (\textit{purple}). Note the different character of the blue curve on the left panel for $\nu=1$.}
\end{figure*}

First, $\rho + P_2$ is approaching zero as $r$ tends to infinity.  What is happening near the horizon $r \rightarrow r_h$ is a more complex issue, because the horizon position cannot be determined analytically. However, Eq. (\ref{enec}) clearly says if $\mu>3$, and the outer horizon is very close to center, NEC can be violated. From Figs. \ref{hornu} and \ref{hormu}, it follows that the outer horizon does not exist when $q\rightarrow q_L$ and sufficiently large $\nu$. In this situation, the NEC is satisfied at the horizon. An example of NEC is shown in figure \ref{fNEC}.

Then is easy to find that  $\rho + P_2$ has no zero in the interval $r\in (r_h,\infty)$. The NEC is fulfilled everywhere in the outer part of spacetime if it is fulfilled at the horizon.

\subsubsection{Weak energy condition}
The weak energy condition is defined as 
\begin{eqnarray}
T_{\mu\nu} u^\mu u^\nu \geq 0,
\end{eqnarray}
where $u^\mu$ is a time-like vector. We can rewrite it and show the physical meaning in the form
\begin{eqnarray}
\rho\geq 0, \quad \rho + P_i \geq 0,\quad i = 1,2,3.
\end{eqnarray}

We examine only $\rho\geq 0$ because the second part was studied in the previous section. Again we start with the limiting cases $r \rightarrow r_h$ and $r\rightarrow \infty$. $\rho$ is greater than zero at the horizon and tends to zero at infinity. 

\begin{figure*}
\includegraphics[width=1\hsize]{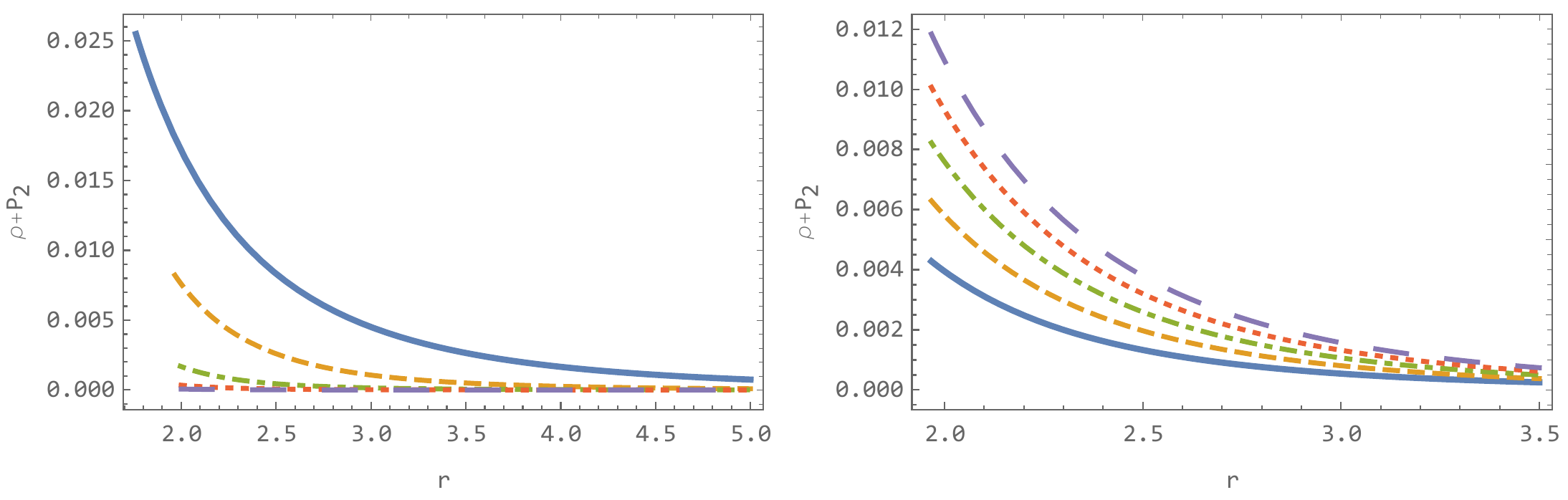}
\caption{\label{fWEC} WEC ($\rho$) of generic black hole spacetime and parameters $q=0.4$ and $\alpha=0.3$. \textit{Left panel} $\mu=4$ and values $\nu$: $\nu=1$ (\textit{blue}), $\nu=2$ (\textit{orange}) and $\nu=4$ (\textit{green}). \textit{Right panel} $\nu=2$ and values $\mu$: $\mu=2$ (\textit{blue}), $\mu=4$ (\textit{orange}) and $\mu=6$ (\textit{green}).}
\end{figure*}

Our condition $\rho\geq 0$ transforms to
\begin{eqnarray}\label{weccon}
\rho = \frac{2 \mu  q^{\nu +3} r^{\mu -3} \left(q^{\nu }+r^{\nu }\right)^{-\frac{\mu +\nu }{\nu }}}{\alpha }\geq 0.
\end{eqnarray}
From Eq. (\ref{weccon}) it immediately follow that it is impossible (for our assumptions) to get negative value for $\rho$. This part ($\rho\geq 0$) of the weak energy condition (WEC) is fulfilled everywhere between horizon and infinity. Examples of $\rho$ profile are shown in Fig.  \ref{fWEC}. The result for the WEC is therefore the same as for the NEC.

\subsubsection{Strong energy condition}

The strong energy condition is defined in the form
\begin{eqnarray}
\Big(T_{\mu\nu} -\frac{1}{2}T g_{\mu\nu}\Big)u^\mu u^\nu \geq 0,
\end{eqnarray}
where $u^\mu$ is a time-like vector. It can be transformed to more physically transparent form
\begin{eqnarray}\label{esec}
\rho + \sum_{i=1}^{3} P_i \geq 0.
\end{eqnarray}
If we use the spacetime symmetry again ($\rho + P_1=0$ and $P_2=P_3$), the condition reads 

\begin{eqnarray}\label{eqsec}
(\nu +1) r^{\nu }-(\mu -1) q^{\nu }\geq 0.
\end{eqnarray}
The limit \mbox{$r\rightarrow \infty$} is zero. But (\ref{eqsec}) can be negative at the horizon if $\mu>1$ and outer horizon is close to the center. The behavior of horizon is seen in Figs. \ref{hornu} and \ref{hormu}.

\begin{figure*}
\includegraphics[width=1\hsize]{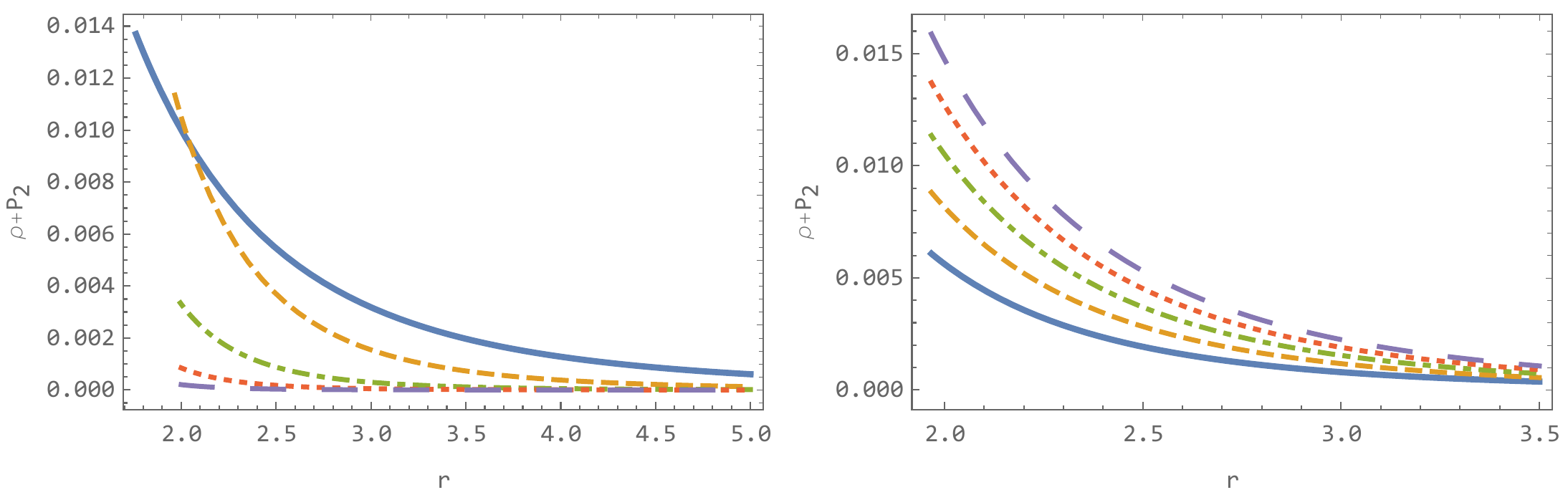}
\caption{\label{fSEC}SEC $P_2$ of generic black hole spacetime and parameters $q=0.4$ and $\alpha=0.3$. \textit{Left panel} $\mu=4$ and values $\nu$: $\nu=1$ (\textit{blue}), $\nu=2$ (\textit{orange}) and $\nu=4$ (\textit{green}). \textit{Right panel} $\nu=2$ and values $\mu$: $\mu=2$ (\textit{blue}), $\mu=4$ (\textit{orange}) and $\mu=6$ (\textit{green}). Note the different character of the blue curve on the left panel for $\nu=1$.}
\end{figure*}

The function $P_2$ has no zero value at interval $r\in(r_h,\infty)$, it results that the strong energy condition (SEC) is fulfilled in outer spacetime  if value at horizon is positive. Fig. \ref{fSEC} presents $P_2$ for several values of parameters $\mu$ and $\nu$.

\section{Neutral test particle motion around generic black hole \label{3circ}}
Now consider the neutral test particle motion around the generic black hole. We distinguish two situations photons and massive particles. We examine the effective potential for both cases and look at the special orbits - unstable circular photon orbit (UCPO) for photons and ISCO for massive particles. The circular geodesics are determined by the local extrema of the effective potential, $\frac{dV_\textrm{eff}}{dr}=0$. These orbits are stable (unstable) for minima (maxima) where $\frac{d^2V_\textrm{eff}}{dr^2}>0$ ( $\frac{d^2V_\textrm{eff}}{dr^2}<0$). The ISCO orbits are given by $\frac{d^2V_\textrm{eff}}{dr^2}=0$, i.e., inflex point. The local extrema of the effective potential cannot be given in an analytical form due to the general form of the lapse function of the generic spacetimes. Therefore, all the calculations are given numerically and results are presented in graphical form. The effective potential for the radial motion of test particle can be obtained from the normalization condition
\begin{eqnarray}\label{normcon}
g_{\mu\nu}u^\mu u^\nu=\delta,
\end{eqnarray}
where $\delta$ is $0$ or $-1$ for photons and for massive particles, respectively. Due to the spacetime symmetries we define constants of motion 
\begin{eqnarray}\label{const}
E&=&-g_{tt}u^t \quad \textrm{and}\quad L=g_{\phi\phi}u^\phi.
\end{eqnarray}
The motion is always fixed to a central plane that can be for simplicity chosen as the equatorial plane ($\theta = \pi/2$). The radial motion is determined by 
\begin{eqnarray}
\Big( \frac{dr}{d\tau}\Big)^2= E^2-V_\textrm{eff},
\end{eqnarray}
where $\tau$ is the particle's proper time and these constants can be identified as energy and angular momentum of the particle at infinity. The effective potential takes the form
\begin{eqnarray}
V_\textrm{eff}=f(r)\Big(\frac{L^2}{r^2}-\delta \Big),
\end{eqnarray}
where $f(r)$ is the lapse function given by Eq. \ref{lapsf}.

\subsection{Orbits of photons}

In this subsection we discuss orbits of photons ($\delta=0$). It is well known that photons do not follow null-geodesics of the generic spacetime as there are coupled to the NED and one had to introduce \textit{effective} metric~\cite{Novello,StuSche19} given by the relation
\begin{eqnarray}\label{eeffmet}
g^{\mu\nu}_{\textrm{eff}}=\mathcal{L}_F g^{\mu\nu}-4\mathcal{L}_{FF}F^\mu_\alpha F^{\alpha\nu},
\end{eqnarray}
where $\mathcal{L}=\mathcal{L}(F)$ is Lagrangian density, $F^{\mu\nu}$ is electromagnetic tensor and $F=F^{\alpha\beta}F_{\alpha\beta}$ is electromagnetic field invariant. Then $\mathcal{L}_F$ denotes differentiation of Lagrangian density $\mathcal{L}$ with respect to invariant $F$ and $\mathcal{L}_{FF}$ is double differentiation with respect to $F$. The optical phenomena in the special cases of Bardeen black hole were for its effective geometry studied in~\cite{StuSche19,ScheStu19}. Here we study the photons motion in the effective geometry of the generic NED black holes.

The photon's orbits can be determined in relation to the impact parameter $b=\frac{L}{E}$ and the effective potential reads~\cite{StuSche19}
\begin{eqnarray}
V_\textrm{eff}=\frac{f(r)}{r^2}\frac{\mathcal{L}_F+2\mathcal{L}_{FF}F}{\mathcal{L}_F}.
\end{eqnarray}
The radius of circular photon orbits is given by $\frac{dV_\textrm{eff}}{dr}=0$, their impact parameter is given by the effective potential at the given radius.

\begin{figure*}
\includegraphics[width=1\hsize]{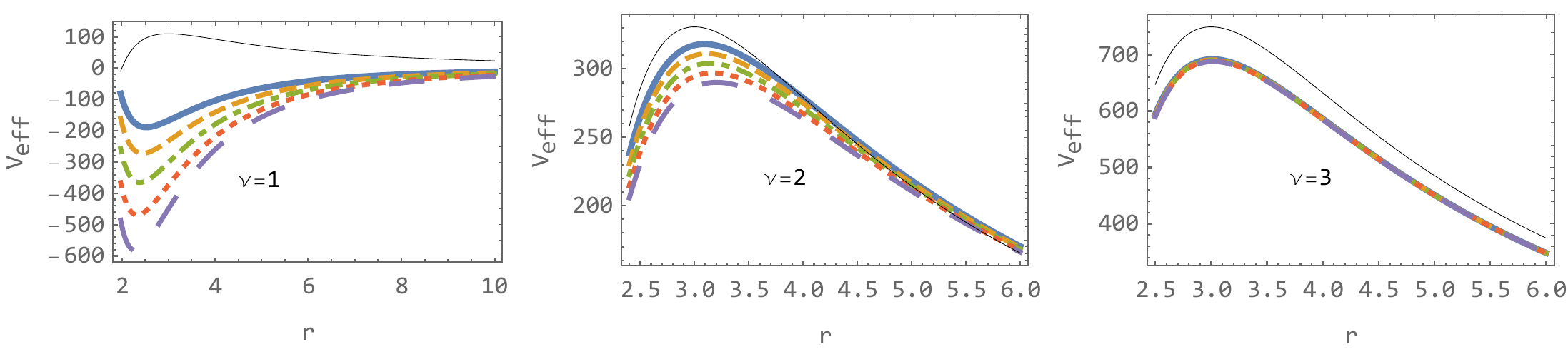}
\caption{\label{ven0}The effective potential of photons described by the effective geometry with parameters $q=0.4$, $\alpha=0.3$, $\nu=1,2,3$ and values $\mu$: $\mu=2$ (\textit{blue}), $\mu=3$ (\textit{orange}), $\mu=4$ (\textit{green}), $\mu=5$ (\textit{red}) and $\mu=6$ (\textit{purple}). Thin black line is Schwarzschild effective potential rescaled from the left by the factor 250, 750 and 1700.}
\end{figure*}

\begin{figure*}
\includegraphics[width=1\hsize]{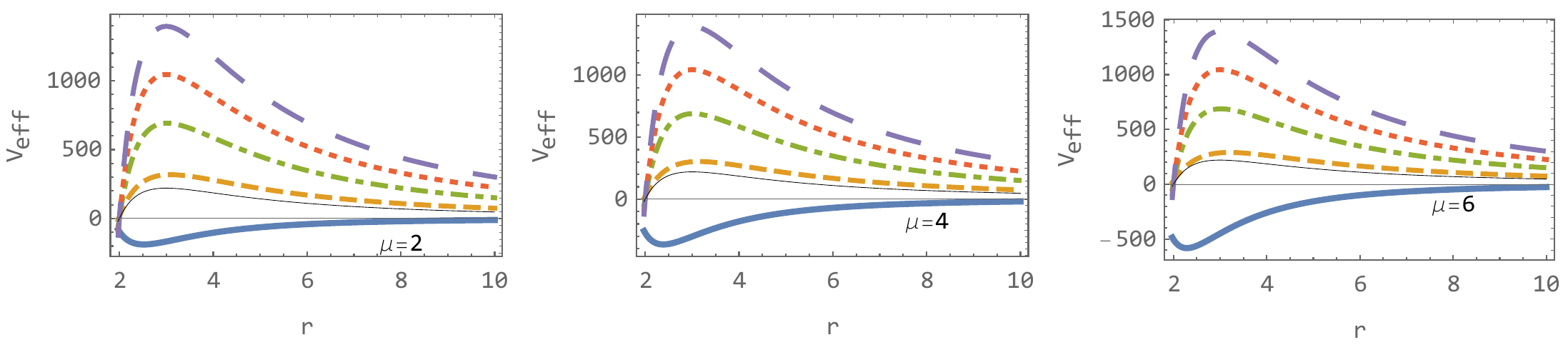}
\caption{\label{vem0}The effective potential of photons described by the effective geometry with parameters $q=0.4$, $\alpha=0.3$, $\mu=2,4,6$ and values $\nu$: $\nu=1$ (\textit{blue}), $\nu=2$ (\textit{orange}), $\nu=3$ (\textit{green}), $\nu=4$ (\textit{red}) and $\nu=5$ (\textit{purple}). Thin black line is Schwarzschild effective potential rescaled by the factor 250.}
\end{figure*}

\begin{figure*}
\includegraphics[width=1\hsize]{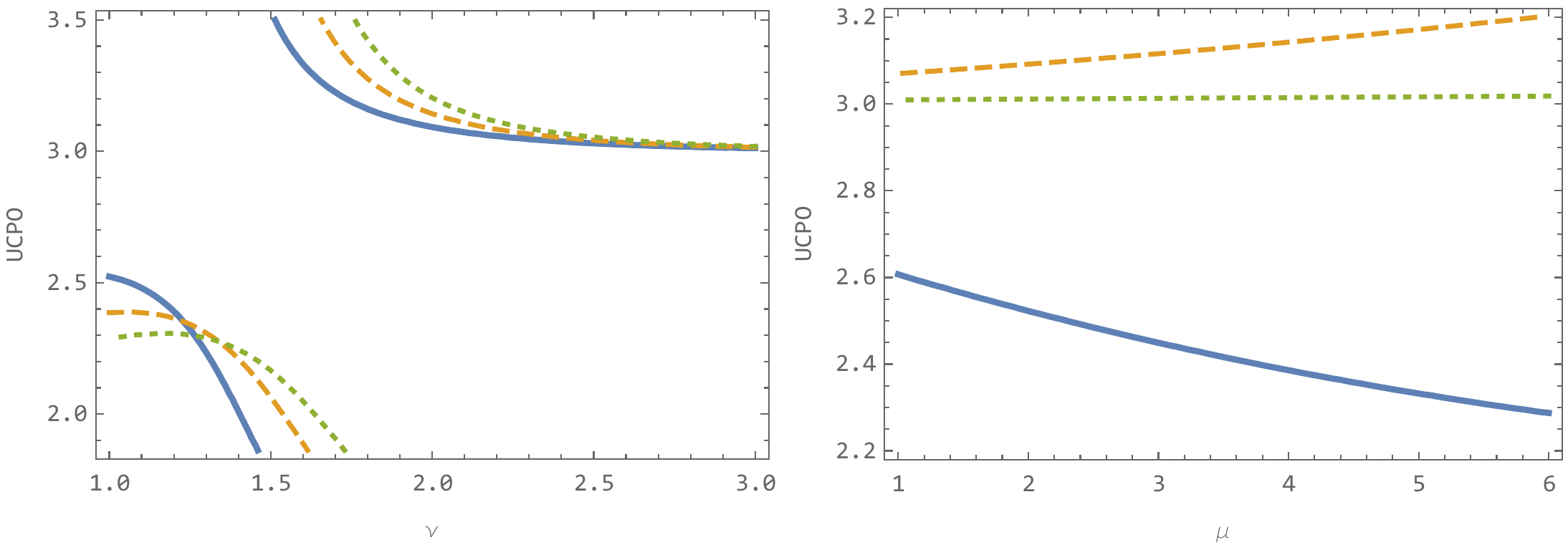}
\caption{\label{rphmn}\textit{Left panel:} The dependence of position of UCPO on parameter $\nu$  and values $\mu$: $\mu=2$ (\textit{blue}), $\mu=4$ (\textit{orange}), $\mu=6$ (\textit{green}). \textit{Right panel:} The dependence of position of UCPO on parameter $\mu$ and values $\nu$: $\nu=1$ (\textit{blue}), $\nu=2$ (\textit{orange}), $\nu=3$ (\textit{green}). All for parameters $q=0.4$, $\alpha=0.3$.}
\end{figure*}

\begin{figure*}
\includegraphics[width=1\hsize]{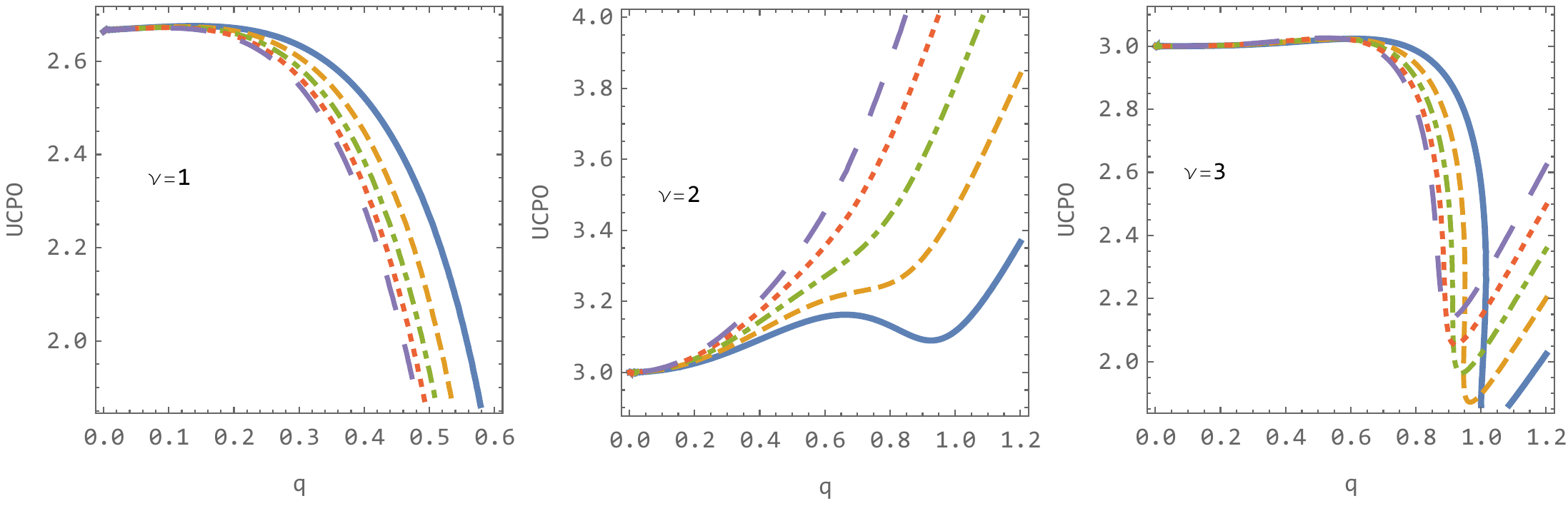}
\caption{\label{rphm} The location of UCPO with parameters $\alpha=0.3$, $\nu=1,2,3$ and values $\mu$: $\mu=2$ (\textit{blue}), $\mu=3$ (\textit{orange}), $\mu=4$ (\textit{green}), $\mu=5$ (\textit{red}) and $\mu=6$ (\textit{purple}).}
\end{figure*}

\begin{figure*}
\includegraphics[width=1\hsize]{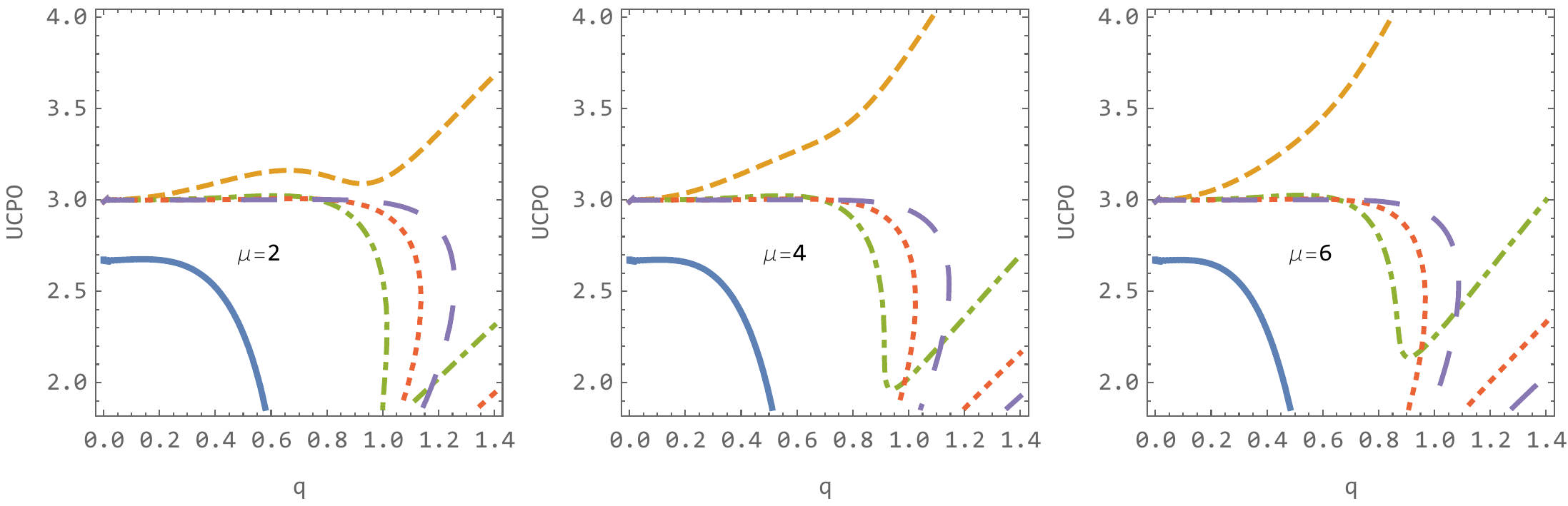}
\caption{\label{rphn}The location of UCPO with parameters $\alpha=0.3$, $\mu=2,4,6$ and values $\nu$: $\nu=1$ (\textit{blue}), $\nu=2$ (\textit{orange}), $\nu=3$ (\textit{green}), $\nu=4$ (\textit{red}) and $\nu=5$ (\textit{purple}).}
\end{figure*}

In Figs. \ref{ven0} and \ref{vem0} typical effective potentials for photons are shown and compared to the Schwarzschild case when the photon motion is governed by the spacetime geometry. The Schwarzschild effective potentials were scaled by constant factors for better resolution - the factors are listed in the description of figures. These factors reflect the large differences of the photons redshift in the effective geometry as compared to the spacetime (Schwarzschild) geometry demonstrated for the first time in the special case of the Bardeen spacetimes in \cite{StuSche19}. The position of UCPO is a little shifted toward the horizon with rising parameter $\mu$. The parameter $\nu$ has a bit different effect. The character of effective potential for small values ($\nu \approx 1$) is very different from that for greater values ($\nu \approx 3$). This peculiar behavior for small $\nu$ causes that in such specific spacetime UCPO becomes stable circular photon orbit, hence there arises a possible window for trapping the photon in potential well in finite area.
This can be seen in Fig. \ref{rphmn} where the influence of parameters $\mu$ and $\nu$ on UCPO is plotted. In the left box, we can identify the change of shape of the effective potential by diverging of UCPO. In this point, the shape is switched and there is no circular photon orbit. How UCPO changes with parameters $\mu$, $\nu$ and $q$ is seen in Figs. \ref{rphm} and \ref{rphn}.

\begin{figure}
\includegraphics[width=1\hsize]{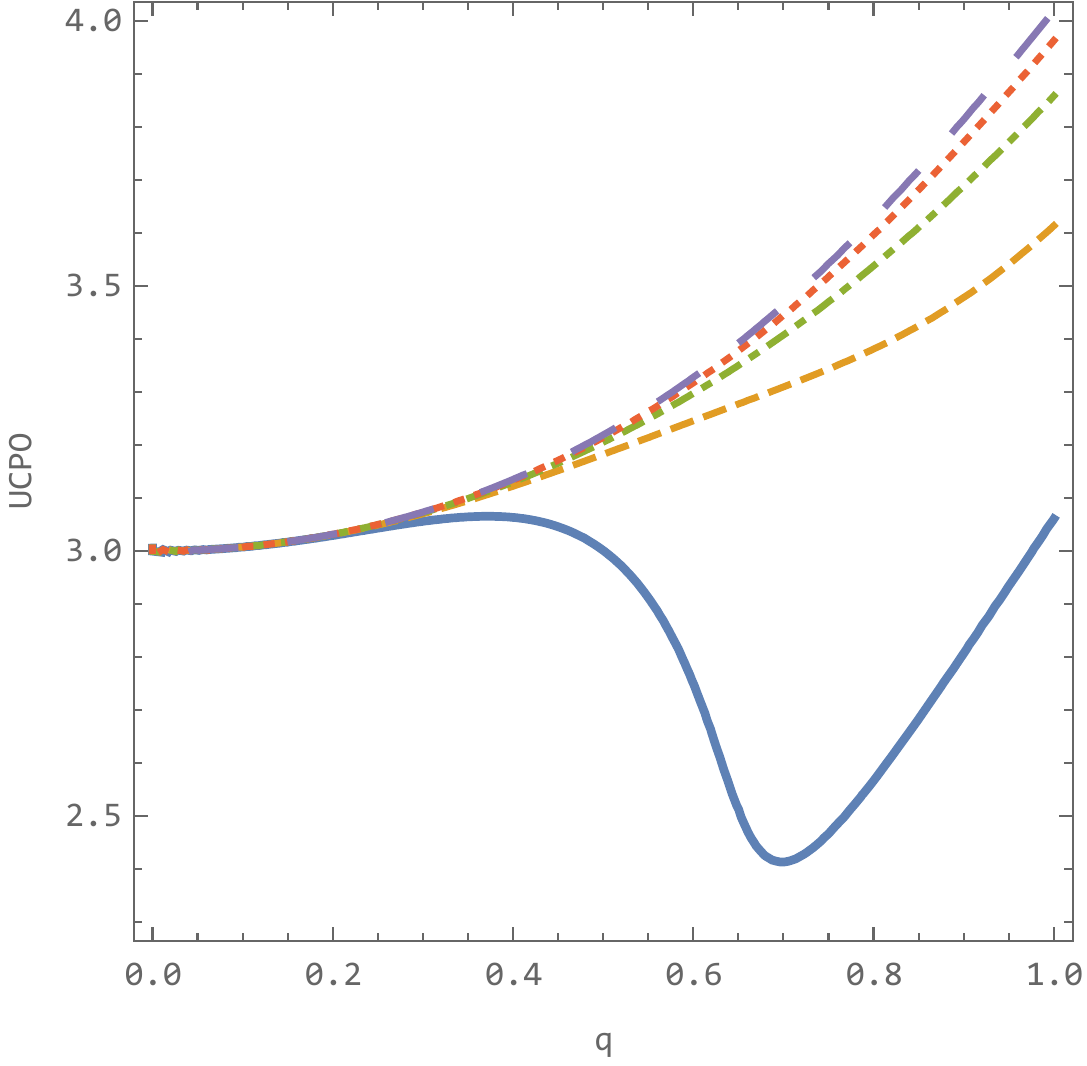}
\caption{\label{rpqa}The location of UCPO for parameters $\mu=3$, $\nu=2$ and values $\alpha$: $\alpha=0.1$ (\textit{blue}), $\alpha=0.3$ (\textit{orange}), $\alpha=0.5$ (\textit{green}), $\alpha=0.7$ (\textit{red}) and $\alpha=0.9$ (\textit{purple}).}
\end{figure}

The influence of the parameter $\alpha$ is shown in the Fig.~\ref{rpqa}.
\subsection{Orbits of massive test particles}
\begin{figure}
\includegraphics[width=1\hsize]{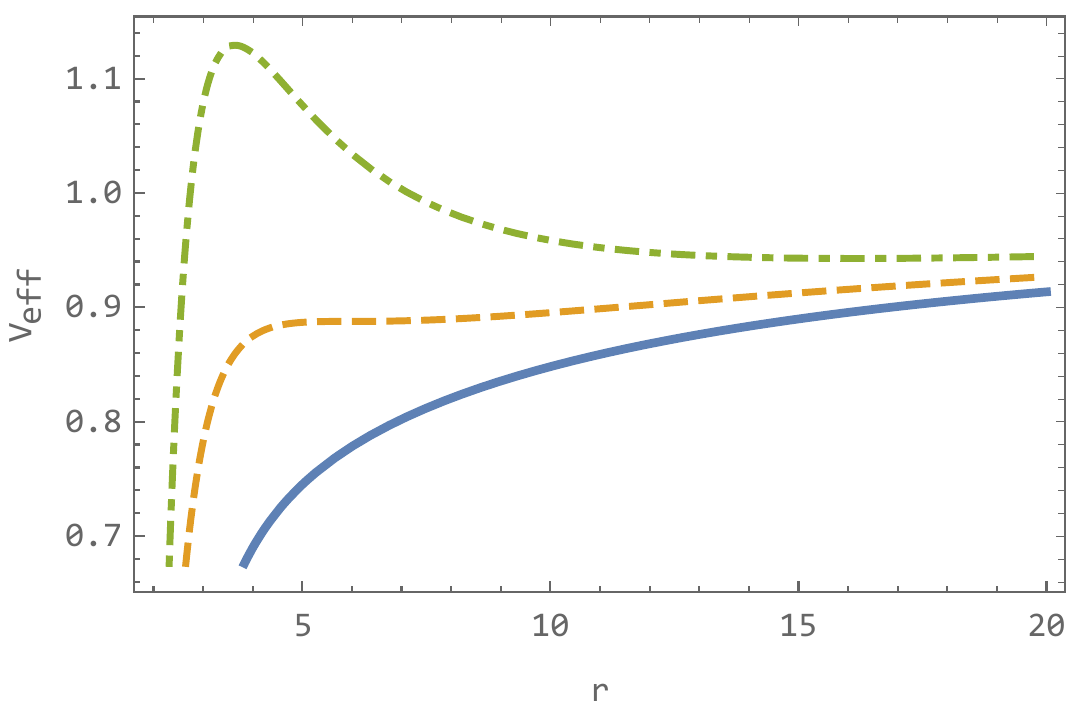}
\caption{\label{vef}The effective potential of massive test particle with parameters $q=0.4$, $\alpha=0.3$, $\mu=3$, $\nu=2$ and values $L$: $L<L_{\textrm{ISCO}}$ (\textit{blue}), $L=L_{\textrm{ISCO}}$ (\textit{orange}), $L>L_{\textrm{ISCO}}$ (\textit{green}), where $L_{\textrm{ISCO}}=3.45$.}
\end{figure}

In this subsection, we examine the behavior of massive test particles ($\delta=-1$). In that case, the effective potential is dependent on angular momentum $L$. There are three possible types of effective potential represented in Fig. \ref{vef}. The most important case is the one with an inflex point that determines ISCO. This situation takes place for specific value of angular momentum, $L_{\textrm{ISCO}}$.

\begin{figure*}
\includegraphics[width=1\hsize]{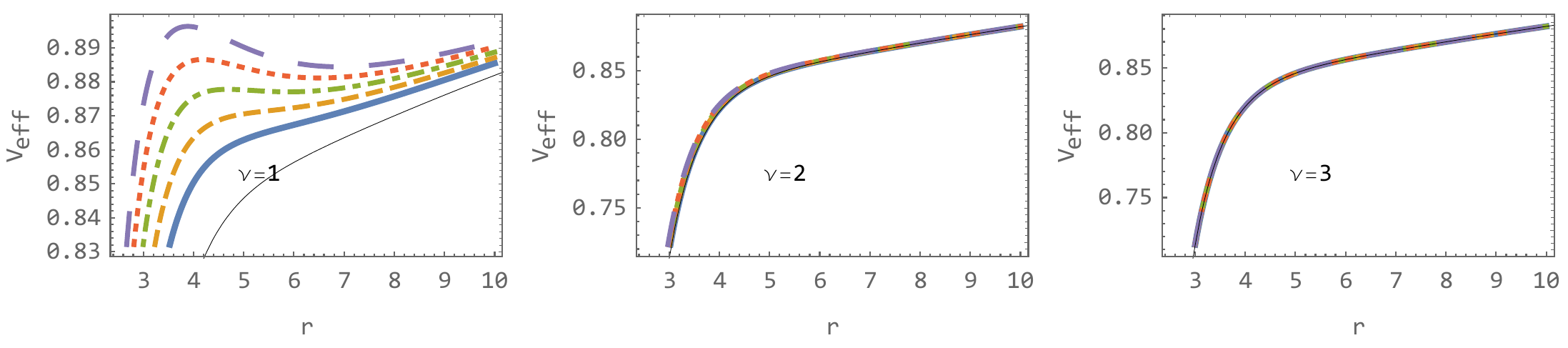}
\caption{\label{ven}The effective potential of massive test particle with parameters $L=3.2$, $q=0.4$, $\alpha=0.3$, $\nu=1,2,3$ and values $\mu$: $\mu=2$ (\textit{blue}), $\mu=3$ (\textit{orange}), $\mu=4$ (\textit{green}), $\mu=5$ (\textit{red}) and $\mu=6$ (\textit{purple}). Thin black line is Schwarzschild effective potential.}
\end{figure*}

\begin{figure*}
\includegraphics[width=1\hsize]{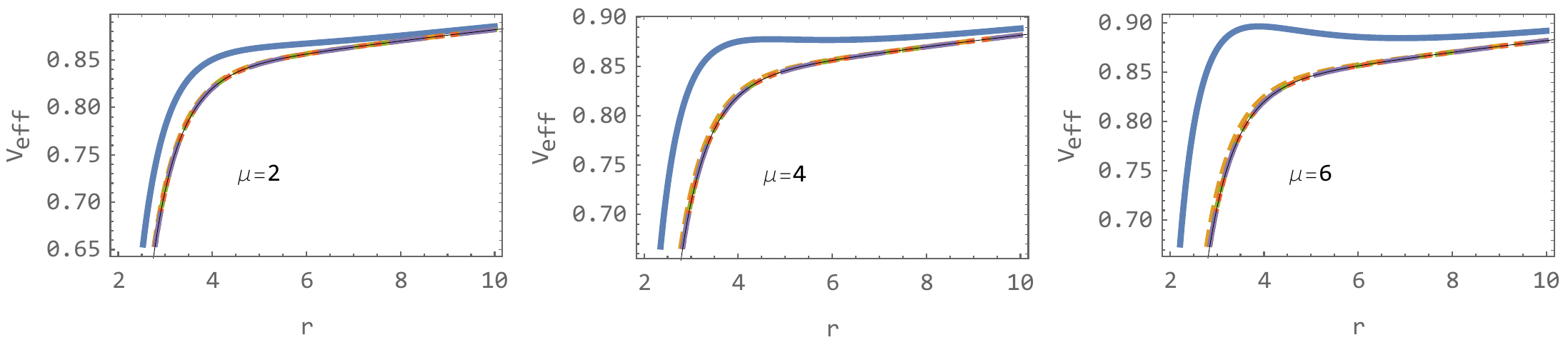}
\caption{\label{vem}The effective potential of massive test particle with parameters $L=3.2$, $q=0.4$, $\alpha=0.3$, $\mu=2,4,6$ and values $\nu$: $\nu=1$ (\textit{blue}), $\nu=2$ (\textit{orange}), $\nu=3$ (\textit{green}), $\nu=4$ (\textit{red}) and $\nu=5$ (\textit{purple}). Thin black line is Schwarzschild effective potential.}
\end{figure*}

\begin{figure}
\includegraphics[width=1\hsize]{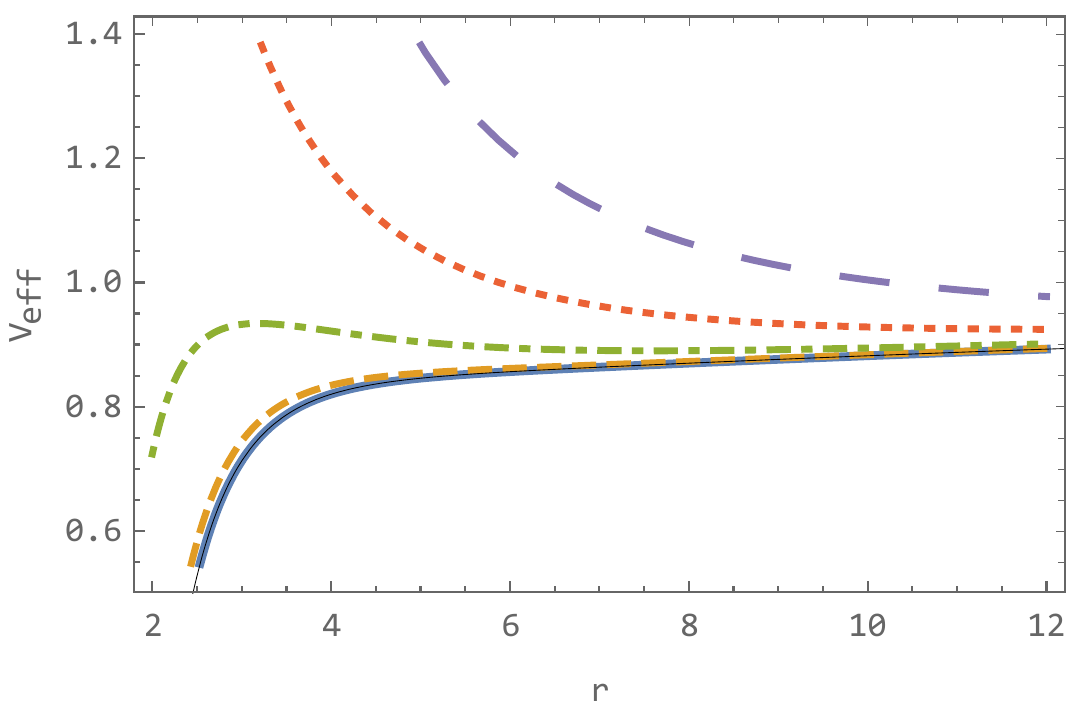}
\caption{\label{veffqa}The effective potential of massive test particle with parameters $L=3.2$, $q=0.4$, $\mu=3$, $\nu=2$ and values $\alpha$: $\alpha=0.1$ (\textit{blue}), $\alpha=0.3$ (\textit{orange}), $\alpha=0.5$ (\textit{green}), $\alpha=0.7$ (\textit{red}), $\alpha=0.9$ (\textit{purple}) and $\alpha=1.1$ (\textit{brown}). Thin black line is Schwarzschild effective potential.}
\end{figure}

The effect of parameters $\mu$, $\nu$ and $\alpha$ on the effective potential is shown in Figs. \ref{ven} - \ref{veffqa}. The role of parameter $\nu$ seems to be similar and even stronger than for photons. Effective potentials with $\nu\gtrsim 1$ and $q\lesssim 0.5$ are very similar to Schwarzschild one, with a weak effect of parameter $\mu$. The effect of parameter $\alpha$ mimics the effect of $L$.

\begin{figure*}
\includegraphics[width=1\hsize]{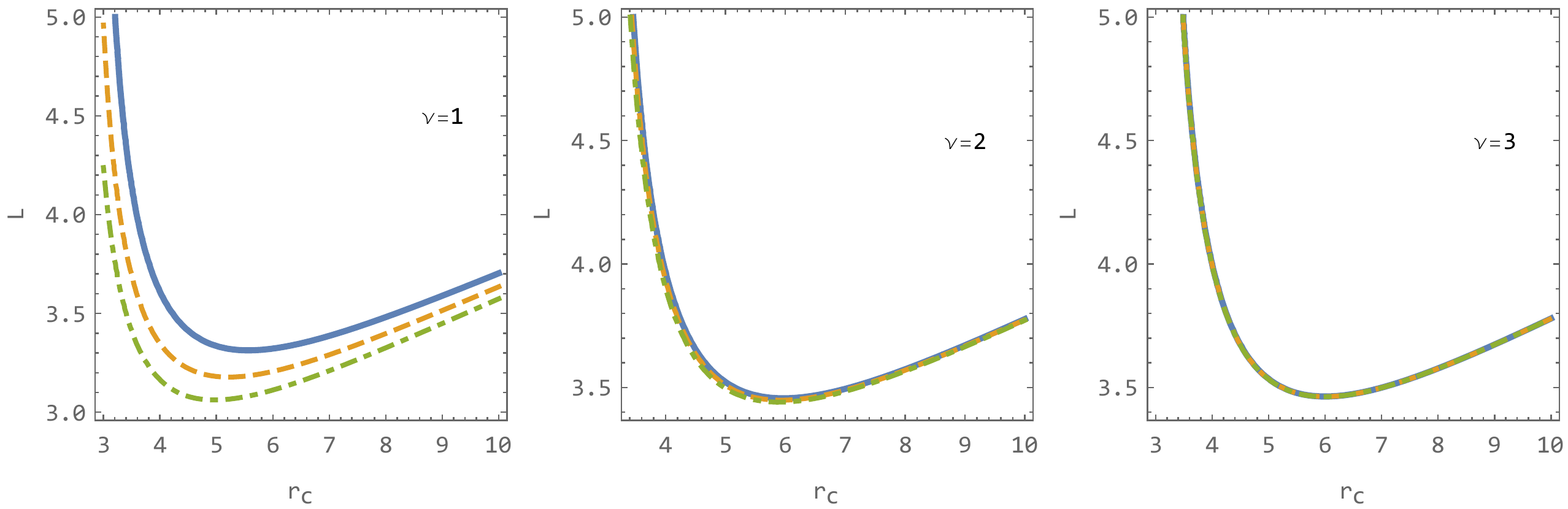}
\caption{\label{rcn}The location of $r_c$ for parameters $q=0.4$, $\alpha=0.3$, $\nu=1,2,3$ and values $\mu$: $\mu=2$ (\textit{blue}), $\mu=4$ (\textit{orange}) (below the green one) and $\mu=6$ (\textit{green}). }
\end{figure*}

\begin{figure*}
\includegraphics[width=1\hsize]{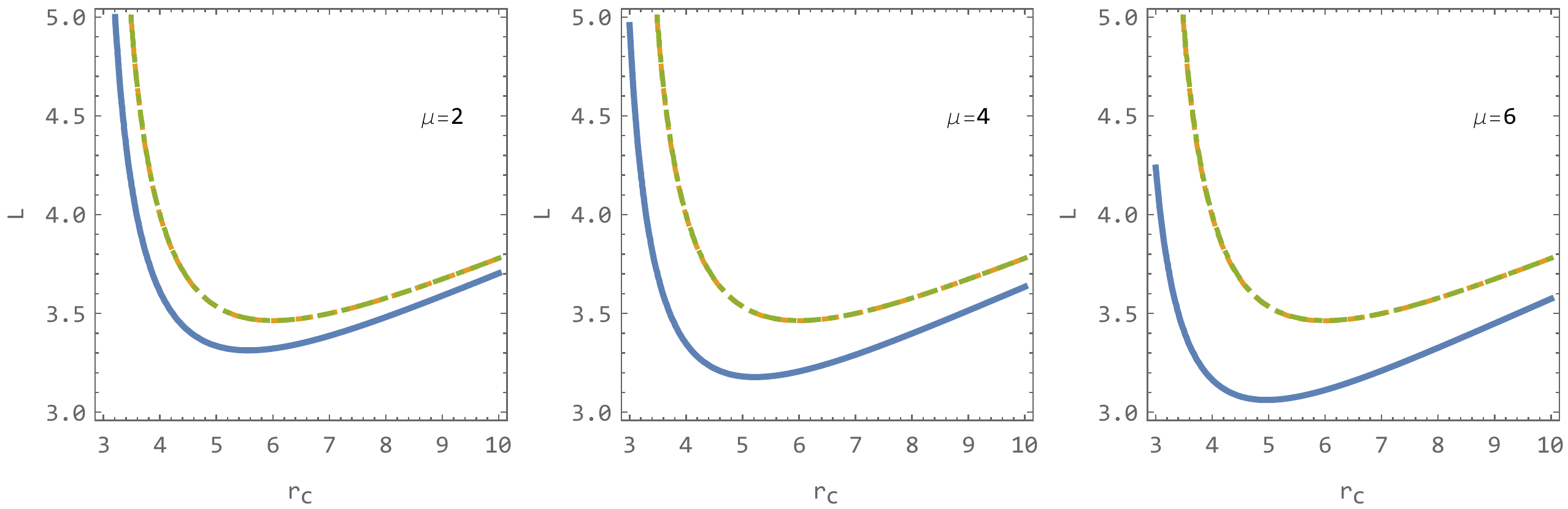}
\caption{\label{rcm}The location of $r_c$ for parameters $q=0.4$, $\alpha=0.3$, $\mu=2,4,6$ and values $\nu$: $\nu=1$ (\textit{blue}), $\nu=3$ (\textit{orange})  (below the green one) and $\nu=5$ (\textit{green}).}
\end{figure*}

The circular geodesics are located at $r_c$, the extrema of the effective potential of a massive test particle that is determined by its specific angular momentum $L$. The position of $r_c$ thus depends on the specific angular momentum $L$. In other words, at a given $r_c$, the specific energy $E$ and the specific angular momentum $L$ are determined by the local extremes of the effective potential. The radial profiles of $L$ are illustrated in Figs. \ref{rcn} and \ref{rcm}. Again one can see that for $\nu \gtrsim 1$ and small values of \mbox{$q\lesssim 0.5$} the curves coincide. At the minimum of these curves ISCO is located. The specific energy of the circular geodesic at a given $r_c$ is determined by the effective potential extrema taken at $r_c$ for given $L$. 

\begin{figure*}
\includegraphics[width=1\hsize]{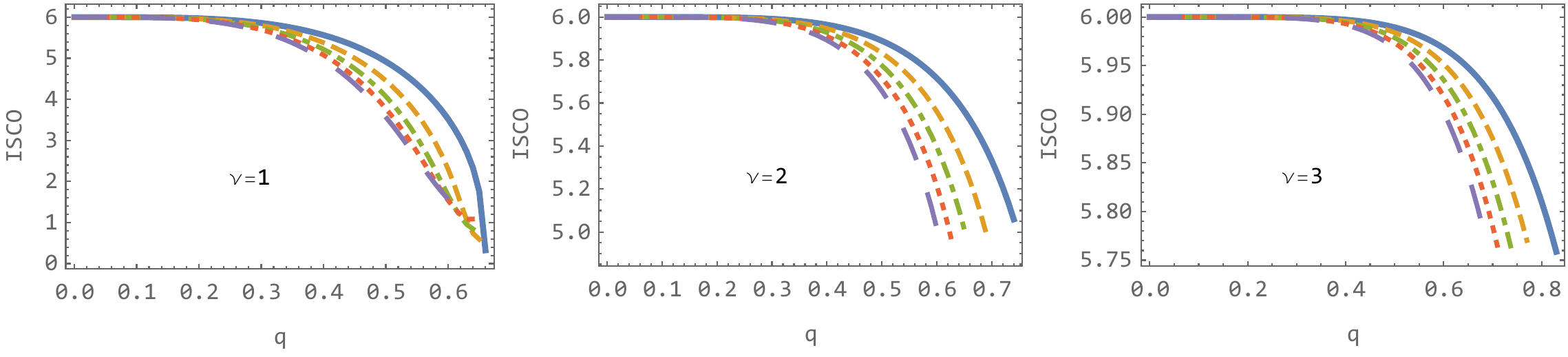}
\caption{\label{iscoqn}The position of ISCO dependent on $q$ for parameters $\alpha=0.3$, $\nu=1,2,3$ and values $\mu$: $\mu=2$ (\textit{blue}), $\mu=3$ (\textit{orange}), $\mu=4$ (\textit{green}), $\mu=5$ (\textit{red}) and $\mu=6$ (\textit{purple}).}
\end{figure*}

\begin{figure*}
\includegraphics[width=1\hsize]{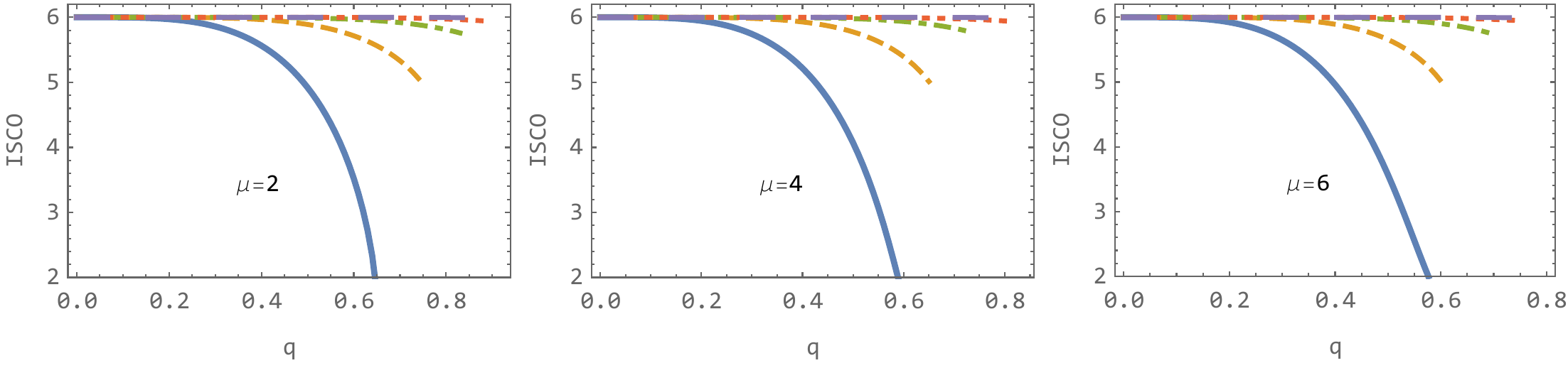}
\caption{\label{iscoqm}The position of ISCO dependent on $q$ for parameters $\alpha=0.3$, $\mu=2,4,6$ and values $\nu$: $\nu=1$ (\textit{blue}), $\nu=2$ (\textit{orange}), $\nu=3$ (\textit{green}), $\nu=4$ (\textit{red}) and $\nu=5$ (\textit{purple}).}
\end{figure*}

ISCO is very important from the astrophysical point of view. Dependence of  ISCO on parameter $q$ for different values of parameters $\mu$ and $\nu$ is shown in Figs. \ref{iscoqn} and \ref{iscoqm}. We can see that this orbit is attracted to the horizon with growing parameter $q$, however, this effect is damped with increasing $\nu$.

\section{Astrophysical application to the SMBH Sgr A*\label{4Mimic}}
The interest in black hole parameters leads us to study accretion processes where ISCO plays a very important role.  ISCO is located at $r=6$ for non-rotating uncharged black hole. For uncharged rotating Kerr black hole ISCO is related to its mass (we set $M=1$) and rotation as~\cite{chandrabook}
\begin{eqnarray}\label{kerris}
r_{\textrm{ISCO}}=3+Z_2\mp\sqrt{(3-Z_1)(3+Z_1+2Z_2)},
\end{eqnarray}
where minus is for pro-grade orbits, which are most interesting for us and plus is for retro-grade orbits,
\begin{eqnarray}
&&Z_1 = 1+\big(1-a^2\big)^{1/3}\Big[\big(1+a\big)^{1/3}+\big(1-a\big)^{1/3} \Big],\\
&&Z_2 = \sqrt{3a^2+Z_1^2}\label{eZ2}\label{kerrz}
\end{eqnarray}
and $a\equiv J/M$ is dimensionless spin parameter of Kerr black hole. ISCO for pro-grade orbits is pushed from non-rotating position ($r=6$) toward the center. The specific energy of co-rotating ISCO comes when formulas for ISCO positions (\ref{kerris}) - (\ref{kerrz}) are used in the formula for specific energy of the circular orbits~\cite{Bardeen73} 
\begin{eqnarray}\label{kerrenis}
E=\frac{r^{3/2}-2r^{1/2}+a}{r^{3/4}(r^{3/2}-3r^{1/2}+2a)^{1/2}}.
\end{eqnarray}

However, we can find out the ISCO position in practice in two ways. Either by direct observation \cite{EHT19} or estimation from the ISCO specific energy \cite{McClint11a,McClint11b,McClintock14,McClin15}.

From the astronomical observations of the black hole binaries (microquasars) it is possible to determine the inner edge of the accretion disk or nearest stable orbits to the center, i.e. the ISCO and estimate the spin parameter $a$. This kind of research has been done for supermassive black hole (SMBH) in the center of Milky way, Sgr A*~\cite{Genzel18}. But one cannot exclude the possibility that some other black hole parameter can mimic the role of the black hole spin. This black hole parameter can be magnetic or electric charge~\cite{Arman1},\cite{Arman2} or alternatively the tidal charge of braneworld black holes \cite{Schee09a,Stuchlik09,Blaschke16}. Here we focus on the possibility to mimic the spin influence by the magnetic charge contained in the generic singular black holes in general relativity combined with NED.

\begin{figure*}
\includegraphics[width=1\hsize]{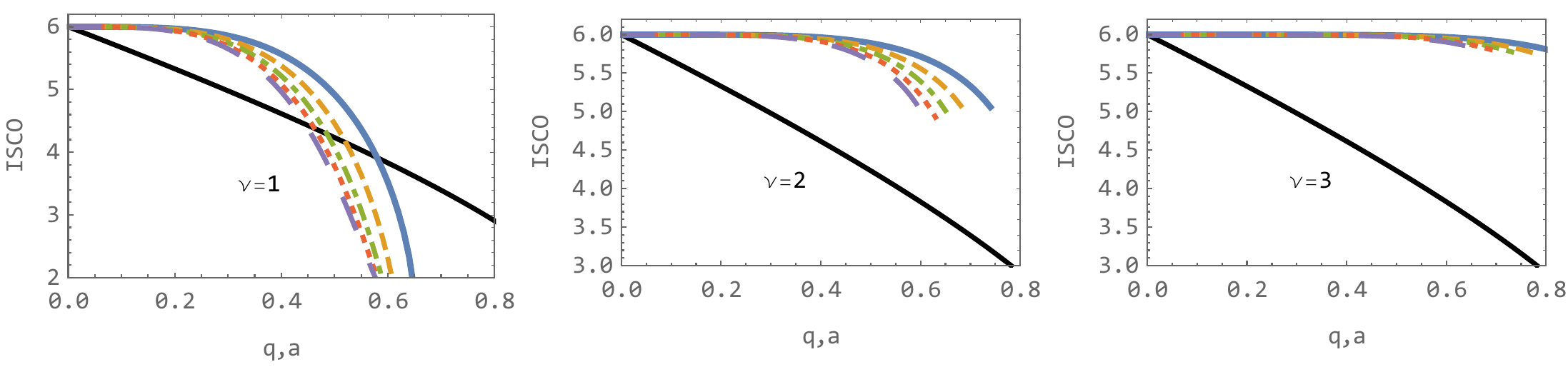}
\caption{\label{iscokern}The position of ISCO dependent on $q$ for parameters $\alpha=0.3$, $\nu=1,2,3$ and values $\mu$: $\mu=2$ (\textit{blue}), $\mu=3$ (\textit{orange}), $\mu=4$ (\textit{green}), $\mu=5$ (\textit{red}) and $\mu=6$ (\textit{purple}) with contrast to position of ISCO for Kerr black hole dependent on parameter $a$ (\textit{black}).}
\end{figure*}

\begin{figure*}
\includegraphics[width=1\hsize]{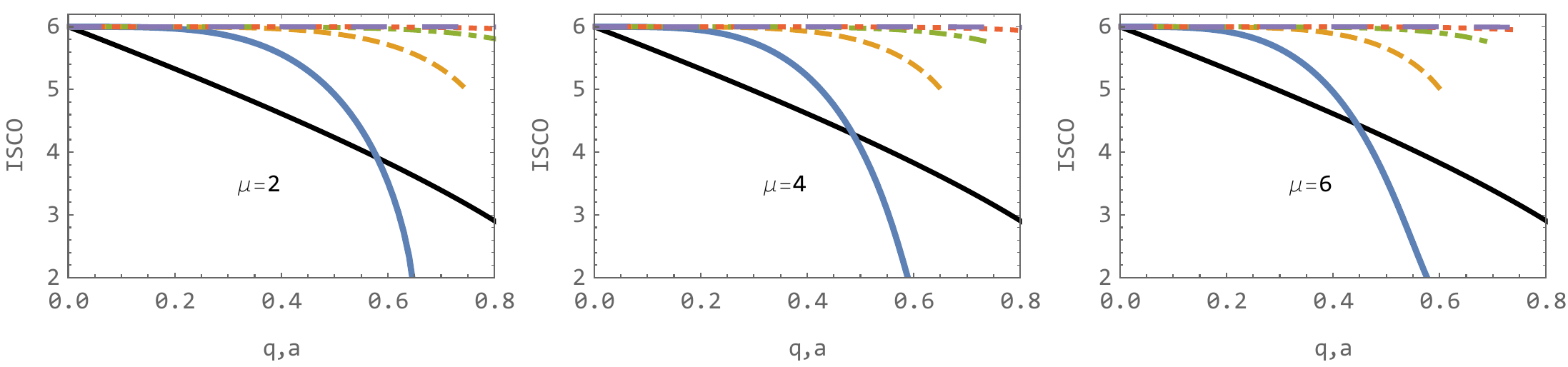}
\caption{\label{iscokerm}The position of ISCO dependent on $q$ for parameters $\alpha=0.3$, $\mu=2,4,6$ and values $\nu$: $\nu=1$ (\textit{blue}), $\nu=2$ (\textit{orange}), $\nu=3$ (\textit{green}), $\nu=4$ (\textit{red}) and $\nu=5$ (\textit{purple}) with contrast to position of ISCO for Kerr black hole dependent on parameter $a$ (\textit{black}).}
\end{figure*}

We estimate here the maximal spin parameter which can be mimicked by an appropriate magnetic charge related to the black hole solutions discussed in our paper. Using previous results presented in Figs. \ref{iscoqn}, \ref{iscoqm} and formulas (\ref{kerris}) - (\ref{eZ2}), one can get how the magnetic parameter can mimic the spin of the black hole in the location of the ISCO.  Figs. \ref{iscokern} and \ref{iscokerm} show ISCO  against charge parameter $q$ in generic black hole spacetime which describes black hole (no naked singularity) and also ISCO for Kerr spacetime against spin parameter $a$. It is obvious that for some combination of parameters can magnetic charge very effectively mimics the rotation creating the degeneracy in black hole parameters. We see that for increasing parameter $\nu$ this phenomenon disappears and the ISCO is close to the Schwarzschild limit.
The maximal values of spin parameter $a$ which can be mimicked by maximal magnetic parameter $q$ for variety of parameters $\mu$ and $\nu$ are summarized in table \ref{tab1}. Values are rounded to two decimals.
\newcolumntype{C}[1]{>{\centering\arraybackslash}p{#1}}
\begin{table}
\begin{tabular}{|C{1.2cm}|C{1.2cm}|C{1.2cm}|C{1.2cm}|C{1.2cm}|C{1.2cm}|}\hline
\multirow{1}{*}{$\mu$}& \multicolumn{5}{c|}{$\nu$}\\ \cline{2-6}
  & 1 & 2 & 3 & 4 & 5 \\   \hline
2 & 1 & 0.28 & 0.07 & 0.02 & 0.01 \\ \hline
3 & 1 & 0.29 & 0.07 & 0.02 & 0.01 \\ \hline
4 & 1 & 0.30 & 0.07 & 0.02 & 0.01 \\ \hline
5 & 1 & 0.31 & 0.07 & 0.02 & 0.01 \\ \hline
6 & 1 & 0.32 & 0.07 & 0.02 & 0.01 \\ \hline
\end{tabular}
\caption{\label{tab1}Tabulated maximal values (rounded to two decimals) of spin parameter $a$ of Kerr spacetime, which can be mimicked by magnetic parameter $q$ of spacetime (\ref{bhmetric}) with mass function (\ref{massf1}) for parameters $\mu=2-6$, $\nu=1-5$ and $\alpha=0.3$ guessed from ISCO position.}
\end{table}

However, estimates based on the ISCO position have to be complemented by estimates of the specific energy of ISCO orbit ($E_{\textrm{ISCO}}$), as this specific energy determines efficiency of the Keplerian accretion and is thus crucial for the black hole parameters in connection with the observational effects \cite{McClint11a,McClint11b,McClintock14,McClin15}. 

\begin{figure*}
	\includegraphics[width=1\hsize]{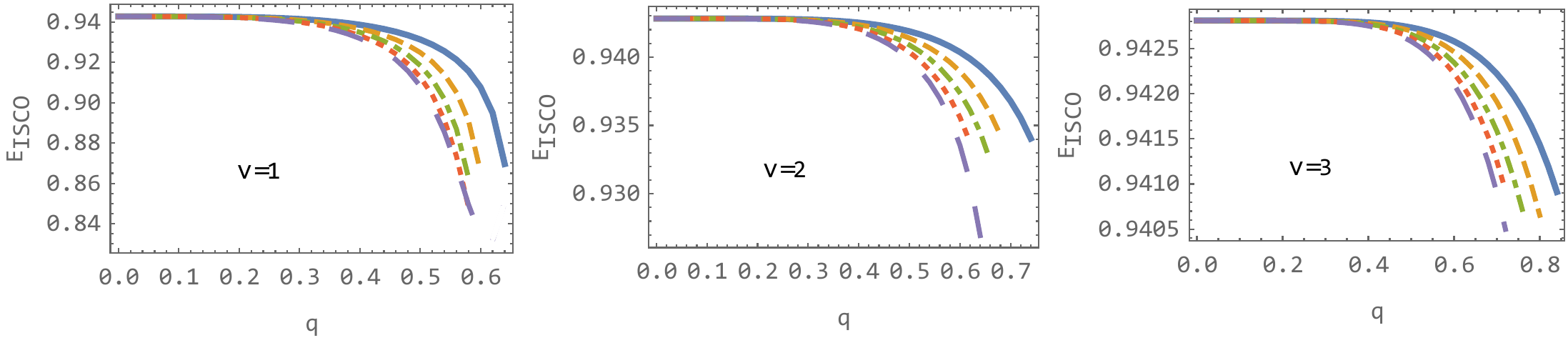}
	\caption{\label{Eiscon}The dependence of specific energy of ISCO on parameter $q$ for parameters $\alpha=0.3$, $\nu=1,2,3$ and values $\mu$: $\mu=2$ (\textit{blue}), $\mu=3$ (\textit{orange}), $\mu=4$ (\textit{green}), $\mu=5$ (\textit{red}) and $\mu=6$ (\textit{purple}).}
\end{figure*}

\begin{figure*}
	\includegraphics[width=1\hsize]{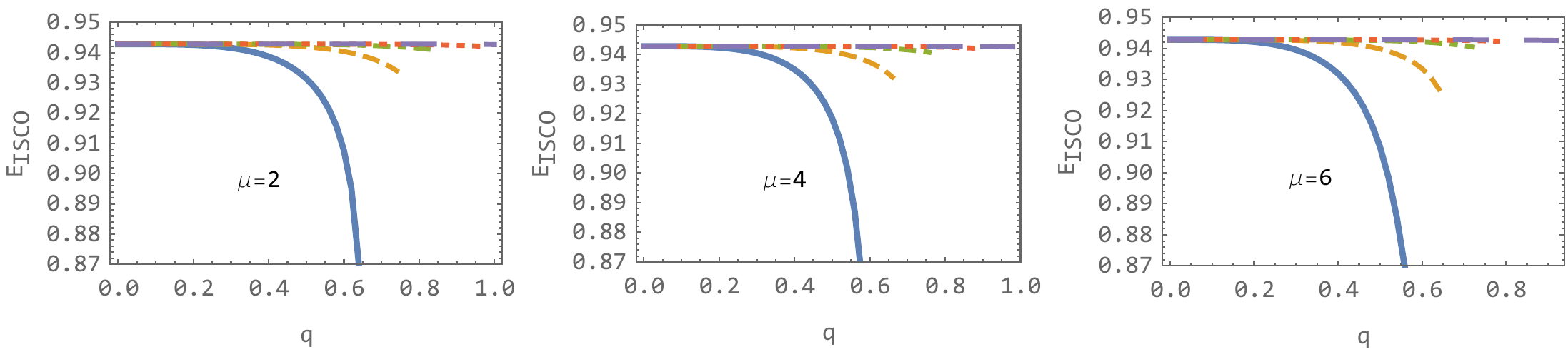}
	\caption{\label{Eiscom}The dependence of specific energy of ISCO on parameter $q$ for parameters $\alpha=0.3$, $\mu=2,4,6$ and values $\nu$: $\nu=1$ (\textit{blue}), $\nu=2$ (\textit{orange}), $\nu=3$ (\textit{green}), $\nu=4$ (\textit{red}) and $\nu=5$ (\textit{purple}).}
\end{figure*}

\begin{figure*}
	\includegraphics[width=1\hsize]{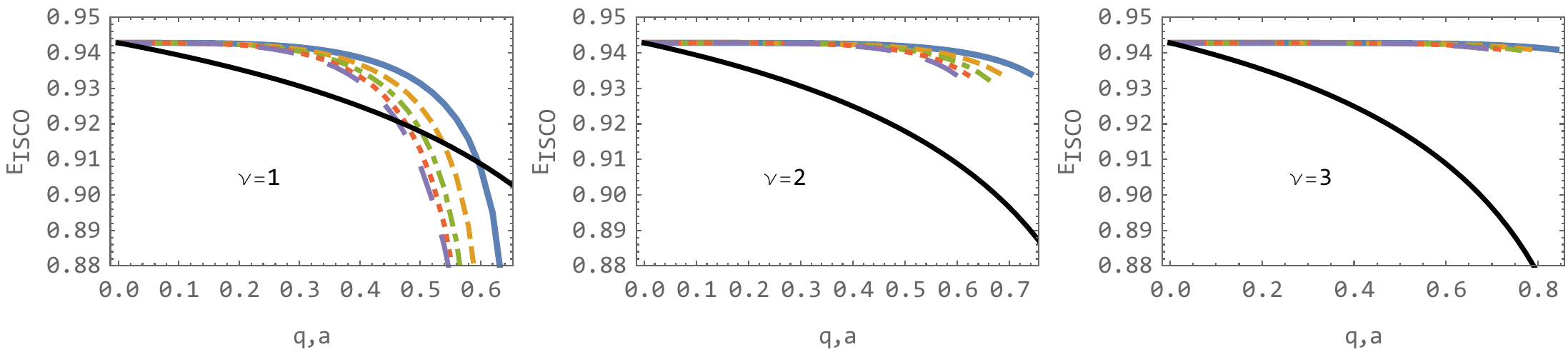}
	\caption{\label{Eiscokern}The dependence of specific energy of ISCO on parameter $q$ for parameters $\alpha=0.3$, $\nu=1,2,3$ and values $\mu$: $\mu=2$ (\textit{blue}), $\mu=3$ (\textit{orange}), $\mu=4$ (\textit{green}), $\mu=5$ (\textit{red}) and $\mu=6$ (\textit{purple}) with contrast to ISCO specific energy for Kerr black hole dependent on parameter $a$ (\textit{black}).}
\end{figure*}

\begin{figure*}
	\includegraphics[width=1\hsize]{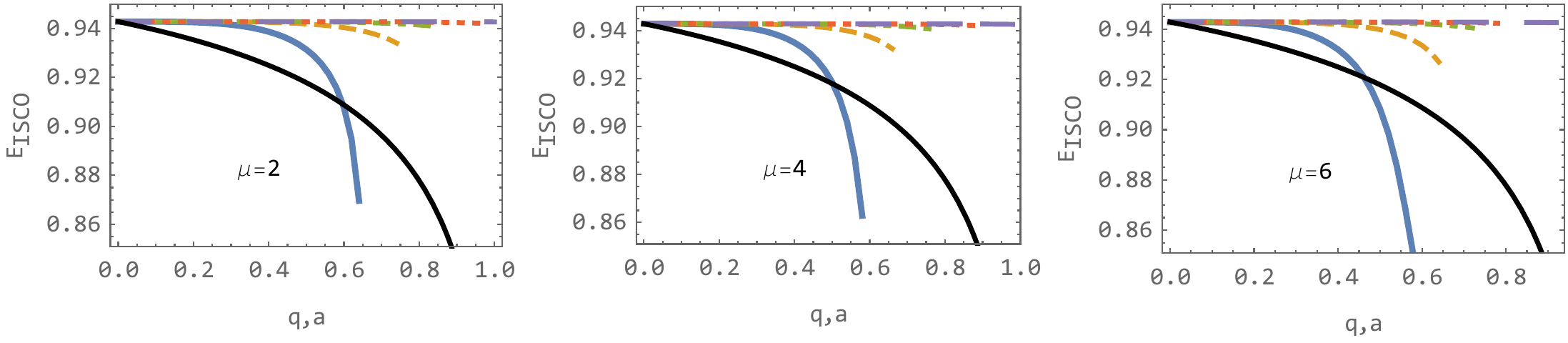}
	\caption{\label{Eiscokerm}The dependence of specific energy of ISCO on parameter $q$ for parameters $\alpha=0.3$, $\mu=2,4,6$ and values $\nu$: $\nu=1$ (\textit{blue}), $\nu=2$ (\textit{orange}), $\nu=3$ (\textit{green}), $\nu=4$ (\textit{red}) and $\nu=5$ (\textit{purple}) with contrast to ISCO specific energy for Kerr black hole dependent on parameter $a$ (\textit{black}).}
\end{figure*}

The dependence of ISCO specific energy on charge parameter $q$ for several values of parameters $\nu$ and $\mu$ is given at figures \ref{Eiscon} and \ref{Eiscom} while its comparison with Kerr ISCO specific energy is illustrated at figures \ref{Eiscokern} and \ref{Eiscokerm}. One can see that the ISCO specific energy is decreasing with increase of the charge parameter $q$, similarly to the Kerr spacetime where the specific ISCO energy decreases with increasing spin parameter $a$. In the NED spacetimes this decrease is significant only for small values of parameter $\nu\approx 1$. With the increasing parameter $\nu$ this phenomenon disappears and the specific energy remains close to the Schwarzschild limit for increasing charge parameter $q$.

\section{Conclusion\label{5Summary}}

In the present work we have investigated properties of the generic singular spacetime (\ref{bhmetric}) with mass function (\ref{massf1}), found as solution of Einstein equation coupled to non-linear electrodynamics. 

We started with existence and position of the horizon. There can exist one, two, three or no horizon. Besides, there always exists the zero-radius horizon for $q=\alpha^{1/3}$. The last possibility indicates naked singularity. We have found out that for any specific values of the parameters $\alpha$, $\mu$ and $\nu$, limiting value of magnetic parameter $q_L$ exists which separates black hole and naked singularity solution. This can be seen in Figs. \ref{hornu} and \ref{hormu}. It can be seen from Fig. \ref{hora} that increasing parameter $\alpha$ leads to increasing location of the outer horizon. It is necessary to point out that when "innermost" horizon reaches the zero, the Eq. (\ref{meqmem}) is satisfied and we have at this point singularity free solution if $\mu\geq 3$. We can find one, two or no horizon in that case.

Then we have tested null, weak and strong energy conditions for the generic spacetime. We have found that, with few exceptions, especially when the outer horizon isn't shifted very close to the center and $\nu$ isn't small ($\nu\lesssim 1$), all conditions, NEC, WEC and SEC are met, as can be seen in Figs. \ref{fNEC} - \ref{fSEC}.

In the section \ref{3circ}, the behavior of uncharged test particles has been investigated. It is divided into two parts, the behavior of the photons - electromagnetic waves and massive test particles.

We have constructed effective potential to describe the main characteristics of the particle motion, Figs. \ref{ven0} and \ref{vem0}. Interesting is that for $\nu \lesssim 1$, the effective potential can be reversed and unstable circular photon orbit becomes stable, leading to possibility of trapped areas and affect in the silhouette of the object~\cite{Stuchlik}. We have localized the position of unstable/stable circular photon orbit. The influence of the spacetime parameters on the photon motion is very complex and is depicted in Figs. \ref{rphmn} - \ref{rpqa}. 

Then we have investigated the effective potentials for massive particles, Figs. \ref{vef} - \ref{veffqa}. In this case the innermost stable circular orbit (ISCO) is very important as it determines efficiency of Keplerian accretion disks \cite{Novikov73}. It was expected that with growing parameter $q$ the ISCO approaching the horizon - as in Maxwellian case - but the main role here has again parameter $\nu,$ as shown in Figs. \ref{iscoqn} and \ref{iscoqm}. The greater value of $\nu$ is the lower shift of the ISCO occur with growing $q$ and ISCO stays close to its Schwarzschild position $r=6$.

The shifting of ISCO toward the horizon with increasing the parameter $q$ is similar to the effect of the rotation in the Kerr spacetime. If we admit the existence of magnetic charge in some objects, for example Sgr A* and assume the spacetime described by the generic singular solutions of general relativity combined with NED, then in same choices of its spacetime parameters and charge, the ISCO postiion (and energy) could be equal to those corresponding to the Kerr spacetime. In this sense the NED charge can mimic the spin parameter of Kerr spacetime. This degeneracy can arise if we estimate spin parameter directly from observation of ISCO position using e.g. GRAVITY or EHT, but also from indirect estimate from ISCO specific energy. The comparison of the shift of ISCO in the generic NED spacetimes with those related to the Kerr spacetime is shown in Figs. \ref{iscokern} and \ref{iscokerm}, and change of the specific energy in Figs. \ref{Eiscokern} and \ref{Eiscokerm}. In Table \ref{tab1} we summarize values of the Kerr spin parameter $a$ which can be mimicked by the magnetic charge of the generic NED black hole solution. The significant effect of mimicking of spin by the magnetic charge disappears with increasing parameter $\nu$ and for values  $\nu \gtrsim	3$ it is impossible completely.

Our results show that the ISCO mimic effect works for both ISCO position and energy in the charge and spin interval $q$, $a$ $\in (0.45 - 0.6)$. 

Further investigation will focus on the motion of electrically charged particles in the field of generic black hole.

\section*{acknowledgements}

We acknowledge the institutional support of Silesian University in Opava and the grant SGS/12/2019.
This research is  supported by Grants No. VA-FA-F-2-008, No. MRB-AN-2019-29  and No. YFA-Ftech-2018-8 of the Uzbekistan Ministry for Innovative Development, and by the
Abdus Salam International Centre for Theoretical
Physics through Grant No.~OEA-NT-01.
This research is partially supported by an
Erasmus+ exchange grant between SU and NUUz. One of the authors (J.V.) was supported by the Czech grant LTC18058. B.A. acknowledges Fudan Fellowship towards his
stay at the Fudan University, Shanghai, China.

\section*{References}

\bibliographystyle{apsrev4-1}  
\bibliography{gravreferences}

\end{document}